\documentclass[aps,floatfix,10pt,showpacs,prd,superscriptaddress]{revtex4-2}
\pdfoutput=1

\usepackage{graphicx}	
\usepackage{siunitx}    
\usepackage{amsmath}	
\usepackage{amssymb}	
\usepackage{xcolor}     
\usepackage{makecell}   
\usepackage{array}
\usepackage{subfiles}   
\usepackage{subcaption} 
\usepackage{enumitem}   
\usepackage{hyperref}   
\usepackage[utf8]{inputenc}
\usepackage{gensymb}
\DeclareUnicodeCharacter{2009}{\,} 
\DeclareUnicodeCharacter{0229}{\c{e}} 

\newcolumntype{P}{>{\centering\arraybackslash}m{3cm}}
\newcolumntype{L}{>{\centering\arraybackslash}m{4.5cm}}

\newcommand{\bham}{
\affiliation{Institute for Gravitational Wave Astronomy 
\& School of Physics and Astronomy, University of 
Birmingham, Birmingham, B15 2TT, United Kingdom}}

\newcommand{\nikhef}{
\affiliation{Dutch National Institute for Subatomic Physics, Nikhef, 1098 XG, Amsterdam, Netherlands}}

\newcommand{\vu}{
\affiliation{Vrije Universiteit Amsterdam, 1081 HV, Amsterdam, Netherlands}}

\newcommand{\MPG}{
\affiliation{Max Planck Institute for Gravitational Physics, Hannover, Germany}}

\newcommand{\leibniz}{
\affiliation{Institut f\"{u}r Gravitationsphysik der Leibniz Universit\"{a}t Hannover, Hannover, Germany}}

\definecolor{ElectricPurple}{HTML}{AA23FF}

\begin{document}

\title{Integration of high-performance compact interferometric sensors in a suspended interferometer}

\author{A Mitchell}
\nikhef\vu

\author{J Lehmann}
\MPG \leibniz

\author{P Koch}
\MPG \leibniz

\author{S J Cooper}
\bham

\author{J van Dongen}
\nikhef \vu

\author{L Prokhorov}
\bham

\author{N A Holland}
\nikhef \vu

\author{M Valentini}
\nikhef \vu

\author{C M Mow-Lowry}
\nikhef \vu


\begin{abstract}
Homodyne Quadrature Interferometers (HoQIs) are compact, low noise and high dynamic range displacement sensors designed for use in gravitational wave observatories. 
Their lower noise compared to the displacement sensors used at present makes them valuable for improving the seismic isolation in current and future detectors.
This paper outlines the progression of this sensor from initial production and benchtop tests to in-vacuum static performance and installation in a gravitational wave detector prototype facility. 
A detailed design description is outlined, including the full signal and optical chain required for implementation in detectors. 
The measured in-vacuum static performance indiacates a noise floor of 3-\num{4e-13}\,m/$\sqrt{\rm{Hz}}$ at 10\,Hz. 
Three HoQIs were installed on the beamsplitter suspension at the AEI 10m prototype.
They measured motion of the intermediate mass across the entire bandwidth measured and showed minimal non-linearities and a good robustness to motion in unmeasured degrees of freedom, both important for practical use in dynamic systems such as seismic isolation. 

\end{abstract}
\maketitle


\section{Introduction}
\label{sec:intro}

The binary black hole collision GW150914, observed by the Laser Interferometer Gravitational-Wave Observatory (LIGO) (with data jointly analyzed by the LIGO Scientific Collaboration and the Virgo collaboration) \cite{Collaboration2015} in 2015, signalled the start of an exciting era in gravitational wave detection \cite{Abbott2016}.
Since then, the LIGO and Virgo collaborations \cite{Acernese2014} have made hundreds of detections of gravitational waves \cite{Abbott2021} \cite{LSC2021}, including a Binary Neutron Star (BNS) detection \cite{Abbott2017} which allowed for multi-messenger astronomy, where both gravitational waves and electromagnetic waves are detected from the same source \cite{Abbott2017a}.
The frequency range current detectors are sensitive to is 20\,Hz-5\,kHz allowing binaries of 7-80 solar mass systems to be detected. 
Extending this range to lower frequencies would allow higher mass and higher redshift binaries to be detected.
As the wave frequency increases throughout the inspiral, earlier detection would be possible, increasing potential for multimessenger astronomy.
This is outlined in the LIGO-LF proposal, which describes many reserch avenues that could push detector limits down to 5\,Hz \cite{Yu2018}.

Current detectors will inevitably reach a peak sensitivity upon which improvements become difficult.
Therefore, future detectors (3G) are being designed to detect a broader frequency range of gravitational waves.
Cosmic explorer and the Einstein telescope will aim to be sensitive down to 5\,Hz and 3\,Hz respectively \cite{Evans2021} \cite{Punturo2010}. 
These detectors may additionally provide greater insight into cosmological questions (such as quantum gravity, dark matter and dark energy), test general relativity predictions and theories about primordial black holes and the formation of supermassive black holes \cite{Maggiore2020}.

The aLIGO design sensitivity at 10\,Hz is $\num{1e-19}\, \rm{m}/\sqrt{\rm{Hz}}$ \cite{Collaboration2015} which is a factor of $10^{10}$ lower than ground motion \cite{Abbott2016}. 
The optics of the interferometer must therefore be isolated from the ground. 
This is done using a combination of active and passive techniques. 
Current active pre-isolation detectors (LIGO) are limited by `global control noise' at low frequencies (10-20\,Hz) \cite{Buikema2020} and fully passive isolation detectors (Virgo) do not have sufficient isolation to reach the performance required below 3Hz for 3G sensitivity, rendering the current approaches inadequate for future detectors. 
Whilst global controls deal with the relative position of the optics in the interferometer, local sensing and control reduces the rms motion of the optics. 
Improved local sensing when combined with damping loops will make the plant used in global controls simpler and more stable leading to lower controls noise \cite{Dongen2022}.
To reach 3G sensitivity, improved local sensors are needed. 
This paper will present a displacement sensor which was first presented in \cite{Cooper2018}, named the HoQI. 
It can be used in local sensing and control of seismic isolation systems, as modelled in \cite{Dongen2022} or as a readout mechanism for other sensors such as the cylindrical rotation sensor (CRS), which measures the tilt of a surface, designed for application in GW detectors \cite{Ross2023}.

There are many different kinds of displacement sensors used in the seismic isolation systems of gravitational wave detectors.
Optical sensors such as the optical sensor and electromagnet actuators (OSEM) and optical levers and electronic sensors such as capacitive position sensors (CPS) and linear variable differential transformers (LVDT) are all employed in current detectors.
Following is a short review of the current displacement sensors in use and the HoQI with key figures of merit in Table \ref{tbl:sensors}.

The CPS used in LIGO is the Microsense 8800 and six sensors measure the relative motion between the stages in the suspension systems \cite{Matichard2015}.
The CPS signals are blended with the seismometer signals to provide a more accurate measurement across the required frequency bandwidth. 

LVDTs are made up of three coils, the middle of which is driven with a sinusoidal signal and connected to the test mass at the mid-point between the other coils. 
A displacement of the test mass induces a sinosoidal signal in the other coils which can be calibrated to a distance measurement \cite{Tariq2002}.
LVDTs are used in Virgo's super attenuator alongside accelerometers to provide a relative position measurement.

Optical levers use a collimated light source, directed at a reflection point on a test mass and a position sensitive photodetector (quadrant photodiode or position sensitive diode).
They are used in Virgo in the angular sensing and control to control the angular position of the suspended mirrors.  

The Birmingham Optical Sensor and Electro-Magnetic actuator (BOSEM) is a specific version of the OSEM which is used primarily in LIGO.
It uses a shadow sensor (LED and flag system) and coil-magnet actuator to sense and correct for position changes in the suspension stages \cite{Carbone2012}, allowing for damping and DC alignment at the top mass in the pendulum chain \cite{Shapiro2012}. 
BOSEMs are also employed on lower masses in the pendulum for global control in the interferometer \cite{Robertson2002,Carbone2012}. 

HoQI is a compact homodyne quadrature interferometer which uses three photodiode outputs and birefringent elements to create a low noise, high dynamic range displacement sensor. 
The sensitivity published in \cite{Cooper2018} is better than all of the currently employed displacement sensors, however, there are multiple sensors under development which claim similar sensitivity \cite{Isleif2016,Gerberding2015,Smetana2022}.
The typical path for a new device to be implemented in detectors is benchtop tests followed by testing in prototype interferometers and based on these results it can then be considered.
For example, the BOSEMs were chosen for use in the aLIGO upgrade due to the mature technology and proven reliability \cite{Carbone2012}.
This paper will demonstrate the maturity of the HoQI through a detailed design review (Section \ref{sec:design}), in-vacuum static testing (Section \ref{sec:static}) and results of an experiment in which three HoQIs were used to simultaneously measure motion in the beamsplitter suspension at the AEI 10m prototype (Section \ref{sec:suspensions}).
The HoQI has out-performed current sensors in multiple locations and under clean environments, showing its reliability and technical readiness, something which the aforementioned new sensors have not yet accomplished.

\begin{table}[h]
    \centering
    \caption{Table comparing different displacement sensors used in gravitational wave detectors. These values come from specific examples of implementations in gravitational wave detectors.}
    \label{tbl:sensors}
    \resizebox{\textwidth}{!}{
    \begin{tabular}{|c|c|c|c|}
    \hline
    \textbf{Sensor} & \textbf{Resolution at 10\,Hz} & \textbf{Limitations}  & \textbf{Calibration} \\ \hline

    \textbf{HoQI} & \num{4e-14}\,m/$\sqrt{\mathrm{Hz}}$ \cite{Cooper2018} & \begin{tabular}[c]{@{}c@{}}Speed $<\pi/4$ phase change per sample \\ (0.26mm/s for 2048 samples/s)  \end{tabular} & Inherent, limited by non-linearity \\ \hline

    \begin{tabular}[c]{@{}c@{}}\textbf{CPS} \\ 1mm \end{tabular} & \num{4e-10}\,m/$\sqrt{\mathrm{Hz}}$ \cite{Lantz2009} & \begin{tabular}[c]{@{}c@{}} $\pm$ 0.5mm \\ maximum, in-line dynamic range \end{tabular} & Geometry and reference-drive-level \\ \hline

    \begin{tabular}[c]{@{}c@{}} \textbf{LVDT} \\ Multi-SAS \end{tabular}& \num{1e-9}\,m/$\sqrt{\mathrm{Hz}}$ \cite{Heijningen2018} & \begin{tabular}[c]{@{}c@{}} $\pm$ 10mm \\ maximum, in-line dynamic range \end{tabular} & Geometry and reference-drive-level \\ \hline

    \begin{tabular}[c]{@{}c@{}}\textbf{Optical Lever} \\ AEI SPI \end{tabular} & \num{1.5e-12}\,rad/$\sqrt{\mathrm{Hz}}$ \cite{Koehlenbeck2023} & 20\,$\mu$rad  & Geometry dependent \\ \hline

    \textbf{BOSEM} & \num{4.5e-11}\,m/$\sqrt{\mathrm{Hz}}$ \cite{Cooper2023} &\begin{tabular}[c]{@{}c@{}} $\pm$ 0.35mm \\ maximum, in-line dynamic range \end{tabular} & Drifts \\ \hline
    \end{tabular}%
 }
\end{table}

\section{Design and Assembly}
\label{sec:design} %

There are several potential applications  for HoQIs within the gravitational-wave community and each application brings its own set of design constraints. 
For suspension applications a compact footprint and mounting design is key as the space provided is limited and the HoQI must interface in a practical and adjustable way to the current suspensions.
HoQIs must also be able to interface with the control and data system used at current gravitational wave detectors. 
The noise performance of HoQI is desired to be good enough for use in the OmniSens project which requires low-frequency, low noise sensors to be able to measure the small motion of the reference mass \cite{MowLowry2019}. 
Some other functional requirements are alignment, ease of use and maintaining high fringe visibility during operation.
There are currently three versions of HoQI in use around the world with slight differences to suit them to their application: CRS HoQI \cite{Ross2023}, suspension HoQI (used for the experiment in Section \ref{sec:suspensions}) and the OmniSens HoQI \cite{MowLowry2019}.
The differences will be discussed in this section alongside design features of HoQI and the reasons behind them. 
It is split into five subsections: Optics, mechanics, electronics, mounting and optical interface and signal and optics chain.

This paper provides a considerably more detailed description of design features than shown in \cite{Cooper2018}, including polarisation optics, optical fibre interface, and the signal processing chain. 
Many design features can be justified through a priori analysis (for example compact design and mounting), but some required empirical study and iterative improvement (for example the optics used in the arms).

A key design feature which will be discussed is the plane mirror HoQI versus the retroreflector HoQI, for which the difference is the optic used in the interferometer arms. 
Both are functional displacement sensors which have different benefits depending on the application. 
The differences come in the ease of alignment, robustness to misalignment and collimation.

\subsection{Optics} \label{sec:optics}

HoQI is, at its core, a Mach-Zehnder interferometer that uses polarisation optics to measure both quadratures simultaneously. 
The technique was first demonstrated in \cite{Bouricius1970} and has seen considerable development since then.

\begin{figure}[ht]
    \centering
    \includegraphics[scale=0.6]{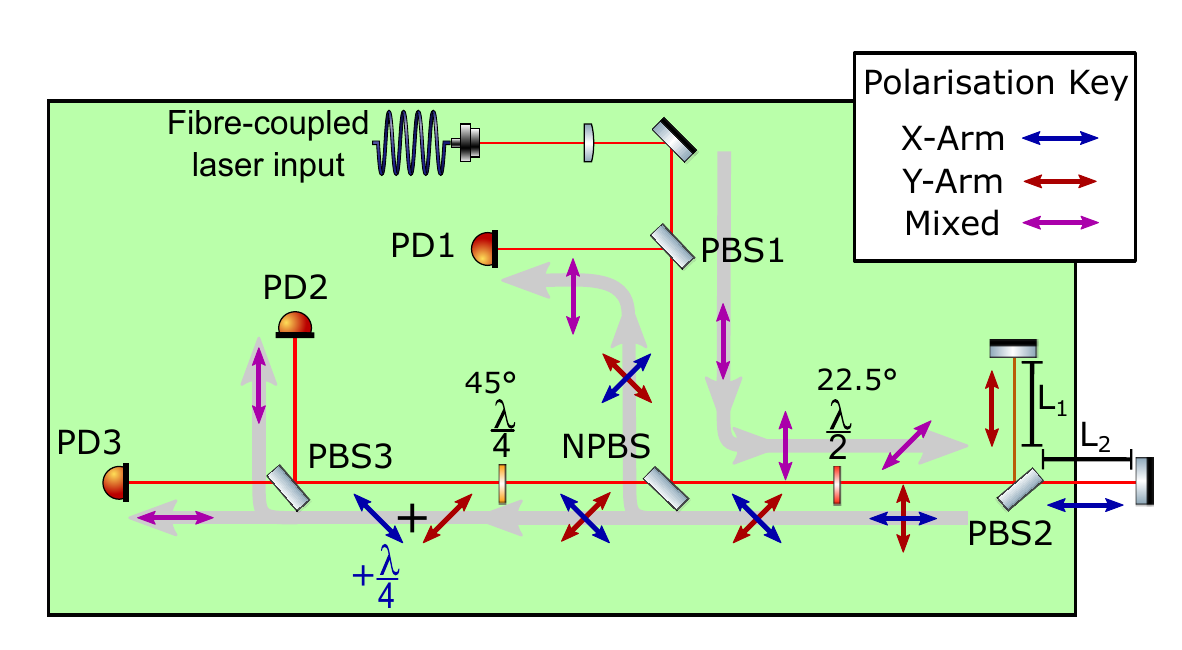}
    \caption{Schematic of the HoQI system. Here PBS is a polarising beamsplitter, NPBS is non-polarising beamsplitter, PD is a photodiode and the half and quarter waveplates are indicated using the $\lambda/2$ and $\lambda/4$, with the angle of the fast axis labelled above them. The direction and polarisation of the beam at each point in the interferometer is indicated by the grey and coloured arrows and the baseplate area is given by the pale green box. }
    \label{fig:HoQISchematic}
\end{figure}

For the optical layout shown in Figure \ref{fig:HoQISchematic}, the optical power at the photodiodes, in the ideal case, will be \cite{Cooper2018}:

\begin{eqnarray}
    P_{\rm PD1}= \frac{P_{\rm in}}{8} (1 + a\sin(\phi)),  \label{eq:PD1} \\
    P_{\rm PD2}= \frac{P_{\rm in}}{8} (1 - a\cos(\phi)),  \label{eq:PD2} \\
    P_{\rm PD3}= \frac{P_{\rm in}}{8} (1 + a\cos(\phi)),  \label{eq:PD3}
\end{eqnarray}

where $P_{in}$ is the input power to the interferometer, $\phi$ is the phase change due to changes in arm length defined by $\phi = \frac{4 \pi}{\lambda}(L_1 - L_2)$, and $a$ is the fringe visibility. 
The redundant information from the third photodiode allows us to reject intensity fluctuations in the input power and fringe-visibility effects by simply subtracting the photodiode signals to create two new quadrature outputs, $Q_1$ and $Q_2$, oscillating about zero and $\pi/2$ out of phase: 

\begin{eqnarray}\label{eq:Q1}
    Q_1 = P_{\rm PD1} - P_{\rm PD2} = \frac{\sqrt{2}a P_{\rm in}}{8} \sin\left(\phi+\frac{\pi}{4}\right), \label{eq:PD1-PD2} \\
    \label{eq:Q2}
    Q_2 = P_{\rm PD1} - P_{\rm PD3} = \frac{\sqrt{2}aP_{\rm in}}{8} \sin\left(\phi-\frac{\pi}{4}\right). \label{eq:PD1-PD3}
\end{eqnarray}

To extract the optical phase, we take the four-quadrant arctangent, which cancels all the common scaling terms and produces an output in radians inherently calibrated to the optical wavelength. 
Unwrapping the phase produces a multi-fringe readout. 
See Section \ref{sec:signal} for more details about this process.

The suppression of intensity noise and inherent calibration of the HoQI readout are substantial advantages. 
There are, however, drawbacks to the polarisation quadrature scheme. 
Principally, any fluctuations in the birefringence over the beam path between the splitting (PBS2) and interferometric recombination (PBS1 and PBS3) are indistinguishable from target motion. 
This is strong motivation for using waveplates with low sensitivity to temperature and angle;
We use true zero-order polymer waveplates (unmounted versions of ThorLabs WPH05ME-1064 and WPQ05ME-1064). 

Two driving design factors in the optical design are minimising non-linearities and maximising the fringe visibility. 
Non-linearities can cause up- and down-conversion in the frequency domain and cross-coupling between different measured degrees of freedom, potentially spoiling performance, see {\it e.g.} Watchi et al. \cite{Watchi2018} and references therein.
It arises from the phase unwrapping of an ellipse, whereas the arctan assumes a circle.
The fringe visibility is a measure of the contrast between dark and bright fringes and though in principle HoQI signal processing eliminates the dependence on it, as in the equations above, reduced fringe visibility has been seen to have a strong effect on noise performance. 
In practice we define a `critical fringe visibility' of 0.5, below which HoQIs show poor performance and increased sensitivity to intensity fluctuations.
Both of these effects are linked and can be reduced by well-aligning the system and ensuring the offset and gain of all photodiode signals is consistent.

Non-linear effects scale with the ellipticity of the Lissajous figure formed by plotting $Q_1$ vs $Q_2$ and with the velocity of the phase change.
The `speed limit' of HoQI is reached when the phase change is greater than $\pi/2$ between consecutive measurements.
The real optical system does not produce exactly equal powers on all three photodiodes, and the phase retardation between the $x$- and $y$-arm beams is not exactly $\lambda/4$.
Both of these effects cause cyclic non-linearities in the optical readout.
Non-linear effects are therefore suppressed by improved waveplate alignment, reduced velocity, and better photodiode gain matching. 
We have not observed significant non-linearity in any well aligned system thus far, especially when the fringe visibility is above the critical fringe visibility. 
Substantial suppression of non-linearities can be achieved via ellipse fitting to the Lissajous figures, and in an intentionally misaligned system a factor of 100 suppression of up-converted signal is demonstrated in Chapter 3 of \cite{Cooper2020}.

The two major drivers of fringe visibility are alignment and polarisation extinction ratio. 
The sensitivity to alignment of the target optic is discussed in Section \ref{sec:optoMech}. 
Currently, we use readily available cube polarising beamsplitters (ThorLabs PBS103) which have very poor extinction ratio, with approximately 20:1 extinction ratio measured in reflection for several units. 
To improve this, plate beamsplitters are being considered for future upgrades.
These could be wedged and custom coated to improve the extinction ratio significantly;
This change would, however, require significant mechanical modifications.

Frequency noise commonly limits noise performance in interferometers. 
To reduce its effect, we match the length of the `reference' and `test' arms. 
In static mirror HoQI tests it is possible to precisely match the arm lengths using three-adjuster mirror mounts. 
Despite careful adjustment, however, we were never able to achieve very deep suppression of frequency noise, measured by modulating the laser frequency and monitoring the equivalent path length change. 
Using Equation \ref{eq:freqNoise} the equivalent arm-length mismatch, $L$, was 100\,$\mu$m. 

\begin{equation}\label{eq:freqNoise}
    \frac{\Delta L}{L} = \frac{\Delta f}{f}
\end{equation}

Since the resolution of our adjustment was at most 10\,$\mu$m, there must be additional coupling from frequency to measured phase, presumably due to scattered light between HoQI optics. 
In retroreflector HoQIs we do not have fine adjustment mounts and this matching is done by hand, meaning we estimate our matching to be within 1mm.
Frequency noise in the retroreflector HoQIs is discussed in Section \ref{sec:static}.

This, in turn, requires that the input laser has small residual frequency fluctuations comparable to those of a high-quality NPRO (we found Innolight Mephisto 500 to be sufficient) or to employ additional frequency stabilisation as presented in \cite{DiFronzo2022}. 
Due to the frequency dependence and working range this also depends on the application;
For some applications which do not require high resolution below 1\,Hz, the laser performance can lead to frequency noise which is acceptable.
For applications with large working range this will be added to the arm-length mismatch potentially increasing frequency noise.

Beam jitter can affect the readout of HoQI due to the size and semiconductor inhomogeneity of the photodiodes.
If there is motion of the input beam, the spot on the photodiodes will move, resulting in differential intensity changes \cite{Kwee2005}. 
To reduce beam jitter and its effects we use fibre collimators with high mechanical stability (Sch\"after + Kirchhoff model 60FC A15-03 and 60FC A4-03 depending on spot size requirement) combined with large area photodiodes (Hamamatsu S2386-8K which are 5.8\,mm square). 
Larger beams also reduce the effect of beam jitter.
The spot size differs between mirror and retroreflector HoQI due to their robustness to misalignment.
The fringe visibility was modelled as a function of misalignment in angle, $\theta$, and translation, $x$ using Taylor expansions of Gaussian beams and can be given by Equation \ref{eq:FVmis} where $\theta_{div}$ is the divergence angle and $x_0$ the beam waist. 

\begin{equation}\label{eq:FVmis}
    a = 1 - \left(\frac{\theta}{\theta_{div}}\right)^2 - \left(\frac{x}{x_{0}}\right)^2 
\end{equation}

Due to the large variation in $\theta$ when a mirror is incorrectly aligned, a large divergence angle beam (smaller beam) is used to compensate for this. 
This may increase the sensitivity to beam jitter so may not be a suitable option if this is not well controlled.

The optical layout shown in Figure \ref{fig:HoQISchematic} has been modified from \cite{Cooper2018} to use the transmission of PBS1 as an input to the interferometer, providing improved polarisation filtering of the input light \cite{ThorlabsUnknown}. 
Due to the polarisation dependence in the interferometer, high extinction ratios are important, especially for the optics prior to the arms, to ensure a clean, single polarisation state in the beam. 
This, combined with the short propagation length of higher-order modes in single-mode fibres and the inherent rejection of intensity fluctuations in the signal processing, means it is difficult to find an acceptable explanation for excess noise based on input optical fibre effects. 
We have empirically found that the quality and strain relief of the upstream fibre, and especially the vacuum feedthrough, is also important.
In practice we use single-mode polarisation-maintaining Panda-980\,nm fibre with FC/APC-PM connections. 
In particular, we use custom UHV-compatible vacuum feedthroughs with connectorised in-vacuum fibre from Diamond GmbH. 
Tests with other vacuum feedthroughs showed excess noise at frequencies between 1 and 10\,Hz without satisfactory explanation.
See Figure 2.7 of Cooper's thesis for an example using a low quality fibre feedthrough, compared to Figure 2.9 which used a better quality feedthrough \cite{Cooper2020}.

\subsection{Mechanics}\label{sec:mechanics}

The mechanics of the HoQIs have been designed to fit curent detector constraints. 
The size on which HoQI could be mounted on the big beamsplitter suspension (BBSS) second mass (M2) tablecloth is 60\,mm x 24\,mm, with the laser height 12\,mm.
Our other design considerations are maximum stability and rigidity for parts fixed to the baseplate.
This compact baseplate along with the lid and optics (mainly) glued to it provide rigidity in the structure.
Flexion of the system would be seen as a position change in the signal, motivating the need for rigidity.

Furthermore, a high thermal mass reduces the DC thermal expansion and filters temperature gradients (which have a similar effect to flexion) by increasing the time it takes to change temperature. 
All metal interfaces in the HoQI connect the same material which reduces the thermal stress between parts due to differing expansion coefficients.

The OmniSens HoQI has an extended reference arm making the baseplate longer at this point. 
This is because the HoQIs are expected to be mounted a bit further away from the test mass and to minimise frequency noise the arm lengths must match. 

The beamsplitters and initial 45\,$\degree$ mirror are glued to the baseplate using vacuum-compatible glue, for example EP30 \cite{EP30} (which has been cleared for use in LIGO).
Other optics (waveplates and retroreflector arm optic) are glued inside peek rings and fixed inside aluminium holders which are bolted to the baseplate. 
Mirrors are held in POLARIS mounts with three adjustment screws which can be bolted to the baseplate.
This allows for rotation of the optics, important to correctly align the waveplate fast axes and the arm optic orientation.
The retroreflector and mirror holders also have the ability to move forwards and backwards in order to match the arm lengths, reducing frequency noise.
The fibre collimator holder has three adjustable feet, allowing the beam position and angle to be adjusted so it hits the centre of the arm optics.
A spherical washer and one bolt to the baseplate allows the collimator angle to be adjustable whilst maintaining a rigid connection to baseplate.

Scattered light would degrade the performance by introducing an optical phase mismatch and therefore frequency noise, so a lid covers it stopping light outside the HoQI from reaching the photodiodes and minimise scattered light in the detector from HoQIs.

\subsection{Electronics}

Whilst the amplifier design has remained mostly unchanged from the original \cite{Hoyland2012}, the photodiodes, cabling and electronics required to bring the photodiode signals to the amplifier has advanced significantly.

Silicon photodiodes (Hammamatsu S2386-8K) are used, with a large photodetection area (as discussed in Section \ref{sec:optics}).
These photodiodes have low capacitance, leakage and dark noise making them a desirable option. 
We use 1064\,nm wavelength light to be able to use silicon photodiodes which allow wavelengths less than 1100\,nm; 
The lower limit is 980\,nm, set by the fibre feedthroughs.  

The three photodiodes and D9 connector are coupled using a kapton flexi circuit, similar to the one used in the BOSEMs. 
This is a flexible sheet of kapton with soldering holes for the photodiode and connector pins which eases the electronics production. 
Previous prototypes of the HoQI flexis showed that there were some strain relief issues, with the connections to the pins becoming loose. 
This has been resolved by reducing bends in the flexi, marking specific bend points and adding kapton reinforcements around the pin holes.

The pre-amplifier input signals are passed through shielded twisted pair (STP) cables which provide good shielding and reduce cost.
The signal chain is designed to use the same BOSEM octopus cables used in LIGO for practicality. 
Despite the capacitance of the photodiode and the cable, the low required bandwidth allows the electronics to be put outside of the vacuum chamber, reducing cost and making the setup easier to build.
The front-end-electronics schematics can be found in \cite{Hoyland2012,Heefner2012,CorporationUnkown}.
It consists of a simple transimpedance amplifier and has the capability to do analogue subtraction to reduce ADC noise and channel usage, however, we have opted to do the signal processing in the DAQ in all presented cases.
This signal chain is very similar to the well-tested one used for the BOSEMs, with some modernisation.

\subsection{Mounting and Optomechanical Interface}
\label{sec:optoMech}

The optomechanical interface in HoQI comes from an optic secured to the target, the test mass. 
As HoQI is designed to be used in a dynamic environment, misalignment of the system can occur when there is motion in degrees of freedom not being measured by the interferometer. 
This motivated the need for two arm optic configurations: one with plane mirrors and one with retroreflectors. 
Plane mirror HoQIs are unaffected by orthogonal translational misalignment within the radius of the mirror but the fringe visibility drops significantly with any angular misalignment.
Conversely, retroreflector HoQIs can withstand greater angular misalignment whilst maintaining good translation robustness.
Previous tests with glass retroreflectors demonstrated low fringe visibility ($a<0.4$) and hence poor noise performance, presumably due to their elliptical output polarisation \cite{Liu1997}. 
Further investigation found that hollow, coated retroreflectors (Newport 50441-0505AU) should better preserve the polarisation state \cite{He2013}.
Static, in-vacuum tests with gold-coated, hollow retroreflectors show an acceptable noise performance whilst maintaining fringe visibility above the critical value despite misalignment, see Section \ref{sec:static} for more information. 
The retroreflector HoQIs have experimentally been found to be easier to align as less adjustment is required to match the beams and maximise fringe visibility, however, residual polarisation effects limit the maximum fringe visibility but this has not been seen to cause problems as long as it is still above the critical value.

\begin{figure}
    \centering
    \includegraphics[scale=0.15]{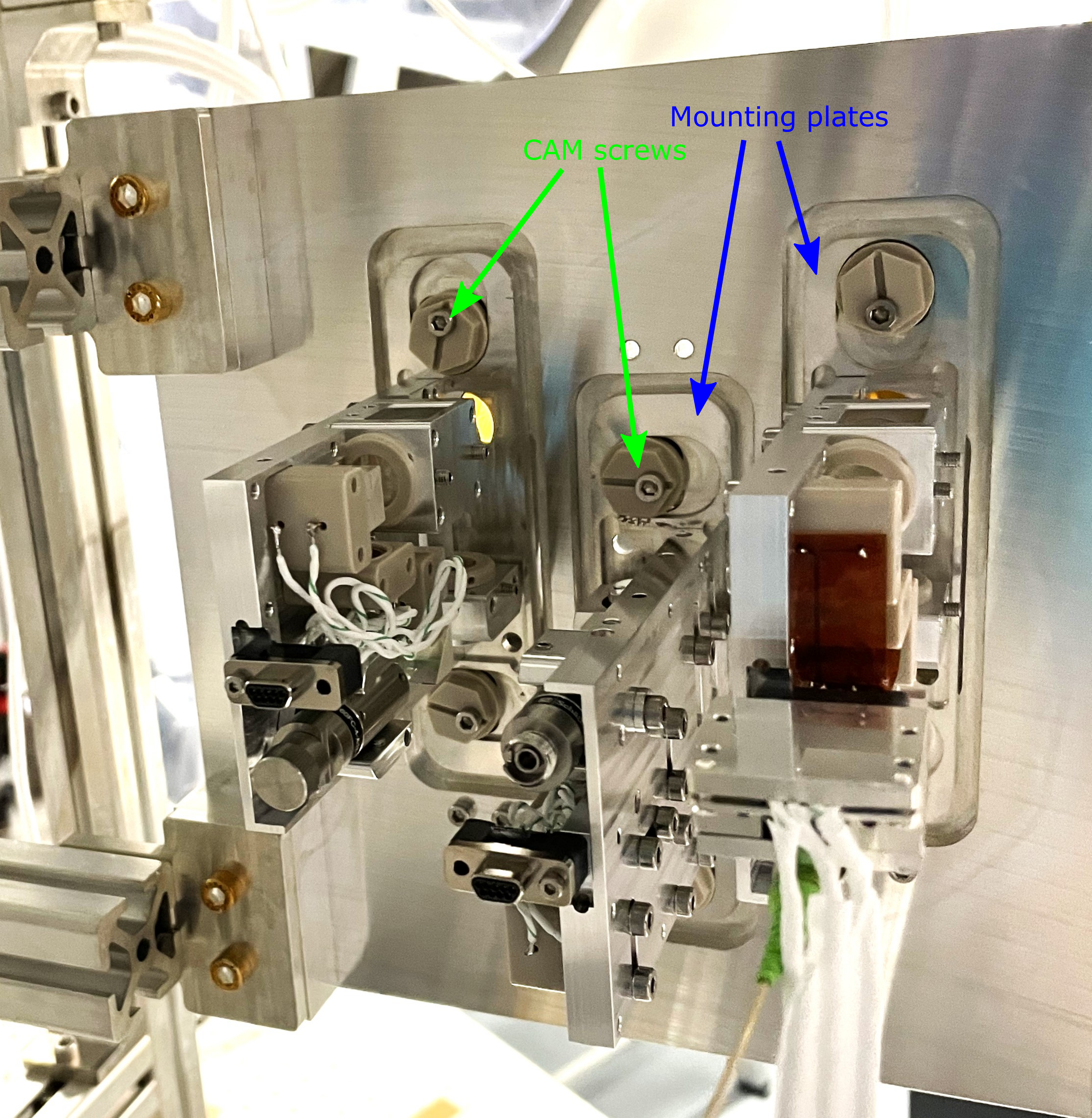}
    \caption{Picture of the HoQI mounted on the AEI suspension system. The CAM screws and mounting plates are labelled. The mounting plate is attached to the tablecloth using the CAM screws which allow for some motion to align the HoQI and they are fixed in place using the M4 screw.}
    \label{fig:AEImounting}
\end{figure}

\begin{figure}
    \centering
    \includegraphics[scale=0.35]{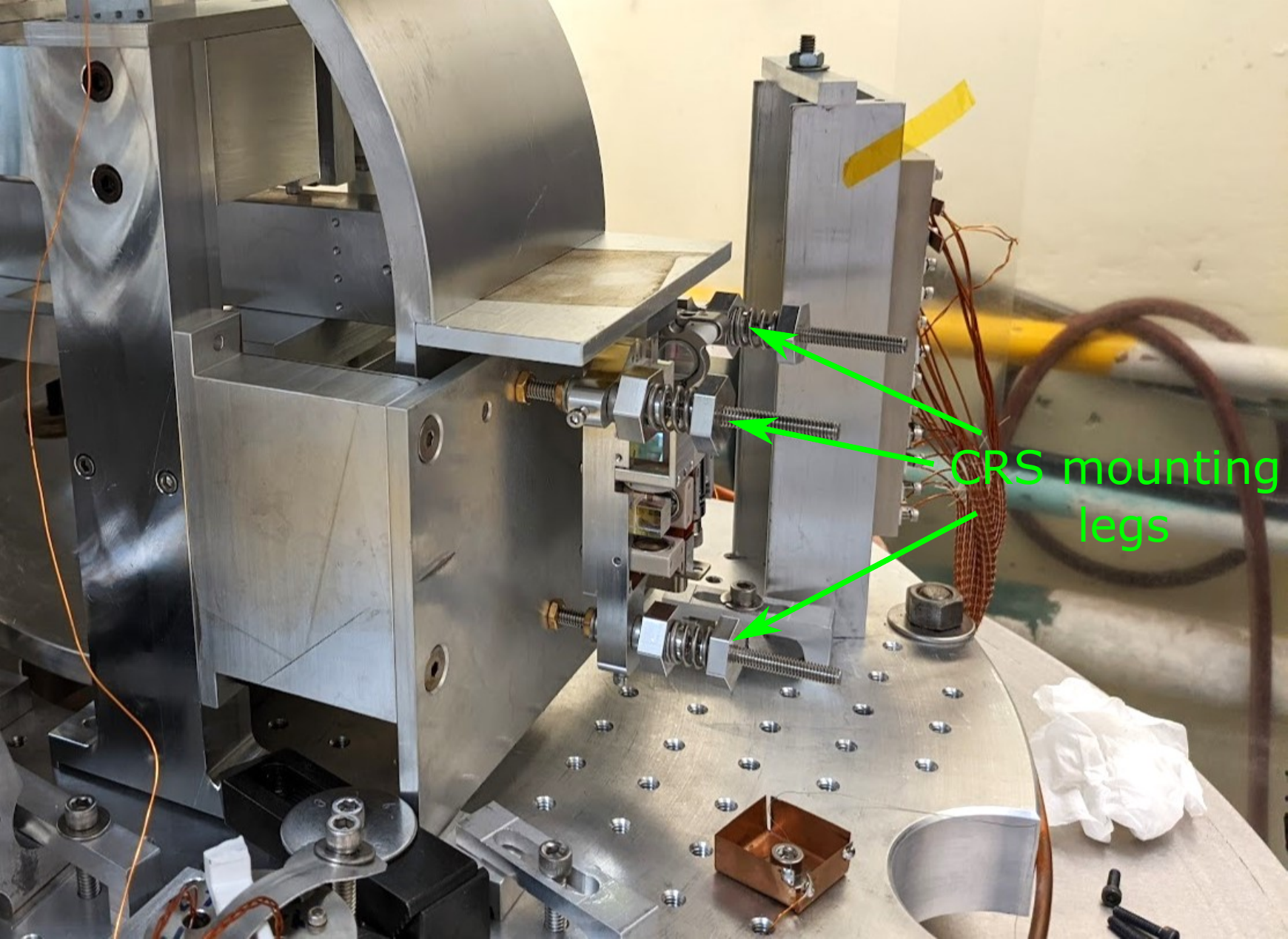}
    \caption{Picture of the HoQI mounted on the CRS. There are three `legs' for each HoQI which have a spring and nut adjustment system to allow for multi-degree of freedom adjustments whilst keeping the HoQI under tension.}
    \label{fig:CRSmounting}
\end{figure}

Interfaces between the HoQI and mounting systems are made to minimise stress but allow for compliance if necessary.
This is done using small areas, similar materials and screws which provide stiffness in the axes necessary but allow for stress relief otherwise. 
The mounting mechanism is the biggest difference between the HoQI variants.
In suspensions and OmniSens, HoQIs are held in mounting plates which can be secured to the `tablecloth' (plate behind the mass) using two adjustment screws \cite{CAM}, allowing for alignment of the HoQI to the test optic vertically and horizontally. 
This can be seen in Figure \ref{fig:AEImounting}.
This gives $\pm$ 2\,mm of range in both directions and is required to center the HoQI on the test arm optic (which generally cannot be adjusted simply) and maximise fringe visibility.
The mounting system was designed by the Rutherford Appleton Laboratory who design and manufacture multiple custom parts for the LIGO collaboration. 
For the CRS a three pronged leg design is used to allow for multi-degree-of-freedom adjustments of the HoQI, this can be seen in Figure \ref{fig:CRSmounting}.

\subsection{Signal and Optics chain}\label{sec:signal}

\begin{figure}
    \centering
    \includegraphics[scale=0.8]{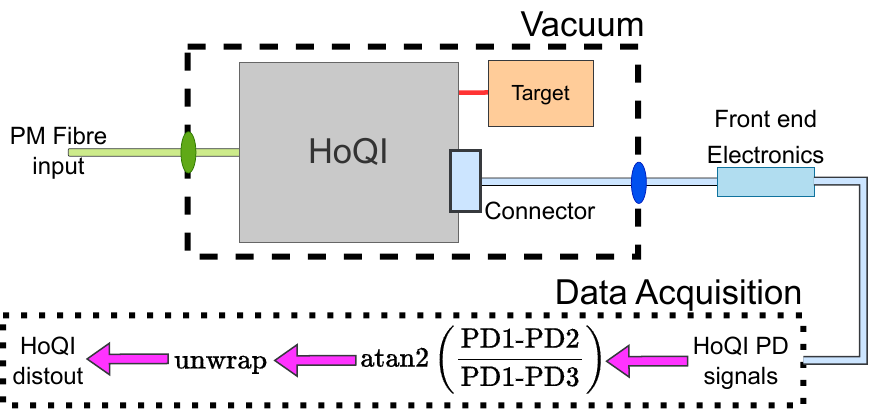}
    \caption{Figure showing a generic signal and optics chain of one HoQI. The green indicates the optics, the blue indicates the signal chain and the pink the data acquisition. An optomechanical interface will be provided by a retroreflector or mirror at the `target'. The oval represents the vacuum feedthroughs for both fibres and electronics. }
    \label{fig:GenSigOp}
\end{figure}

For HoQI to be functional in a system it needs an optical input, optomechanical interface with the target, electrical output and data acquisition.
A schematic of the generic HoQI signal and optics chain is shown in Figure \ref{fig:GenSigOp}.

This shows the flow from the optical chain (PM fibre input, vacuum feedthrough and collimator) to the optomechanical interface (optic on the test mass) followed by the signal chain from the HoQI to vacuum feedthrough (photodiodes and connector) and the subsequent signal processing in the front-end electronics and the data acquisition (DAQ) (using the arctangent and unwrap functions). 
The functions used in the signal processing are standard ones, present in all gravitational wave detector data acquisition systems. 
The `atan2' function does the arctangent which is needed following Equation \ref{eq:Q1} and \ref{eq:Q2} after the division of $Q_1$ by $Q_2$. 
We use this instead of `atan' as this gives us an output in the range $-\pi$ to $\pi$ instead of $-\pi/2$ to $\pi/2$.
This returns $\phi$ the phase difference between the arms, which subsequently is turned into a distance measurement, `distout', using the `unwrap' function.

\section{Static measurements}
\label{sec:static} %

Static measurements allow us to probe the peak performance of a displacement sensor by taking measurements of a static test mass. 
In HoQIs this requires the test arm optic to be mounted onto the baseplate so the motion between arms is minimised. 
Previous prototypes were found to have peak out-of-vacuum sensitivity of \num{2e-14}\,m/$\sqrt{\rm{Hz}}$ at 70\,Hz and \num{7e-11}\,m/$\sqrt{\rm{Hz}}$ at 10\,mHz \cite{Cooper2018}.
Static in-vacuum tests were conducted using the new HoQI prototypes to confirm that they were functional and low noise and to find the in-vacuum noise floor.
With the HoQI in vacuum the noise floor below 0.5\,Hz should be reduced due to the reduction of air currents and temperature fluctuations \cite{Cooper2018}.

\begin{figure}[h]
    \centering
    \includegraphics[scale=0.1]{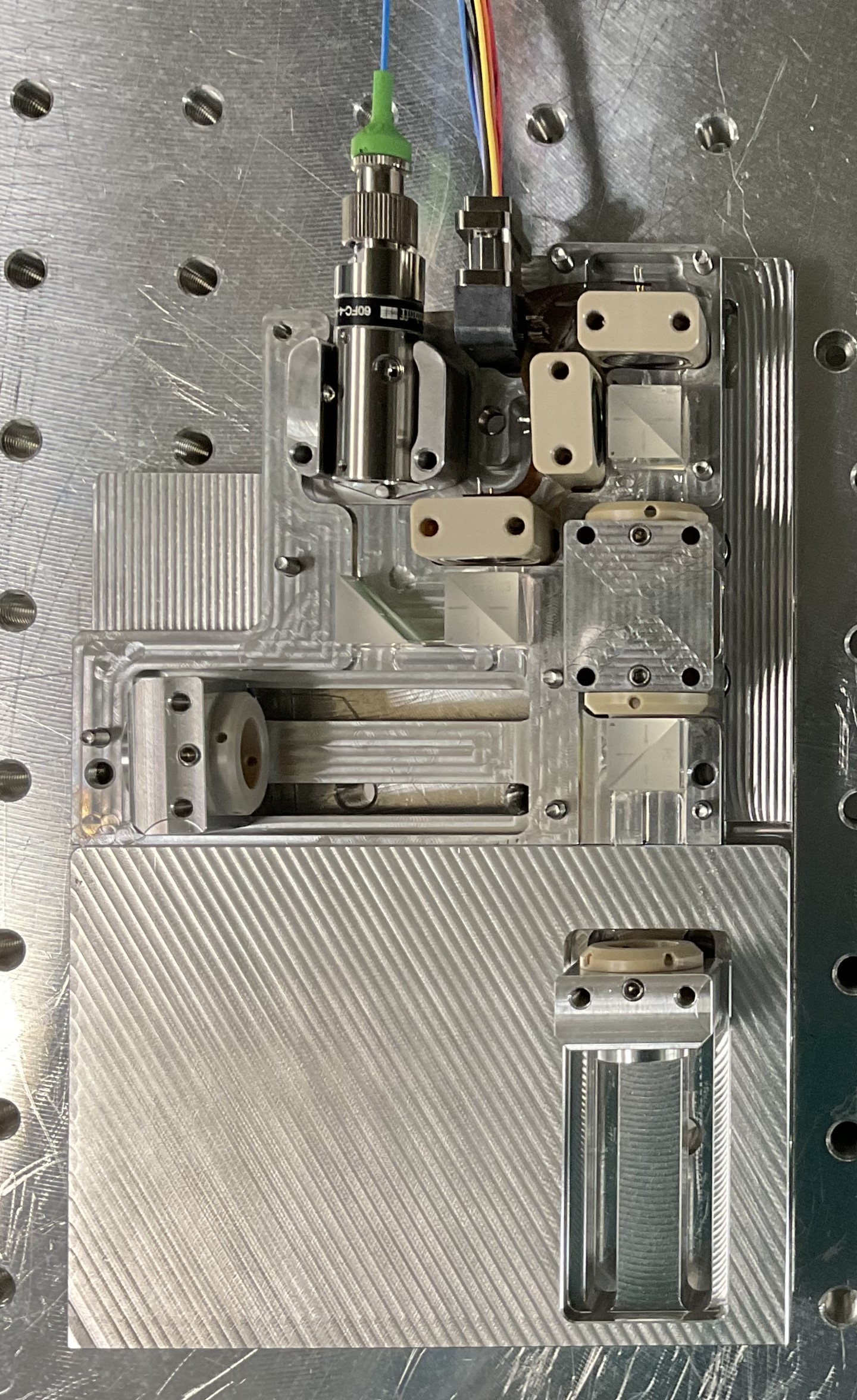}
    \caption{One static HoQI. The extension plate used goes underneath the regular baseplate (see top left and right side of HoQI in picture) and then to the side to create the second arm (see bottom of picture). It is attached to the baseplate using the screw holes usually used for the lid of the HoQI. The fibre and electronics connections can be seen in the top of this photo.}
    \label{fig:HoQIsetup1}
\end{figure}

Static HoQIs were made using extension plates which were screwed to the baseplate of a HoQI from underneath, providing a rigid connection to minimise relative motion between plates. 
Figure \ref{fig:HoQIsetup1} is a picture of a singular static HoQI. 
The HoQI mechanics used has slight variations to those used for the CRS or suspension measurements, however, this should not affect the measurement as all optics and internal arm lengths were consistent.

\begin{figure}[h]
    \centering
    \includegraphics[scale=0.3]{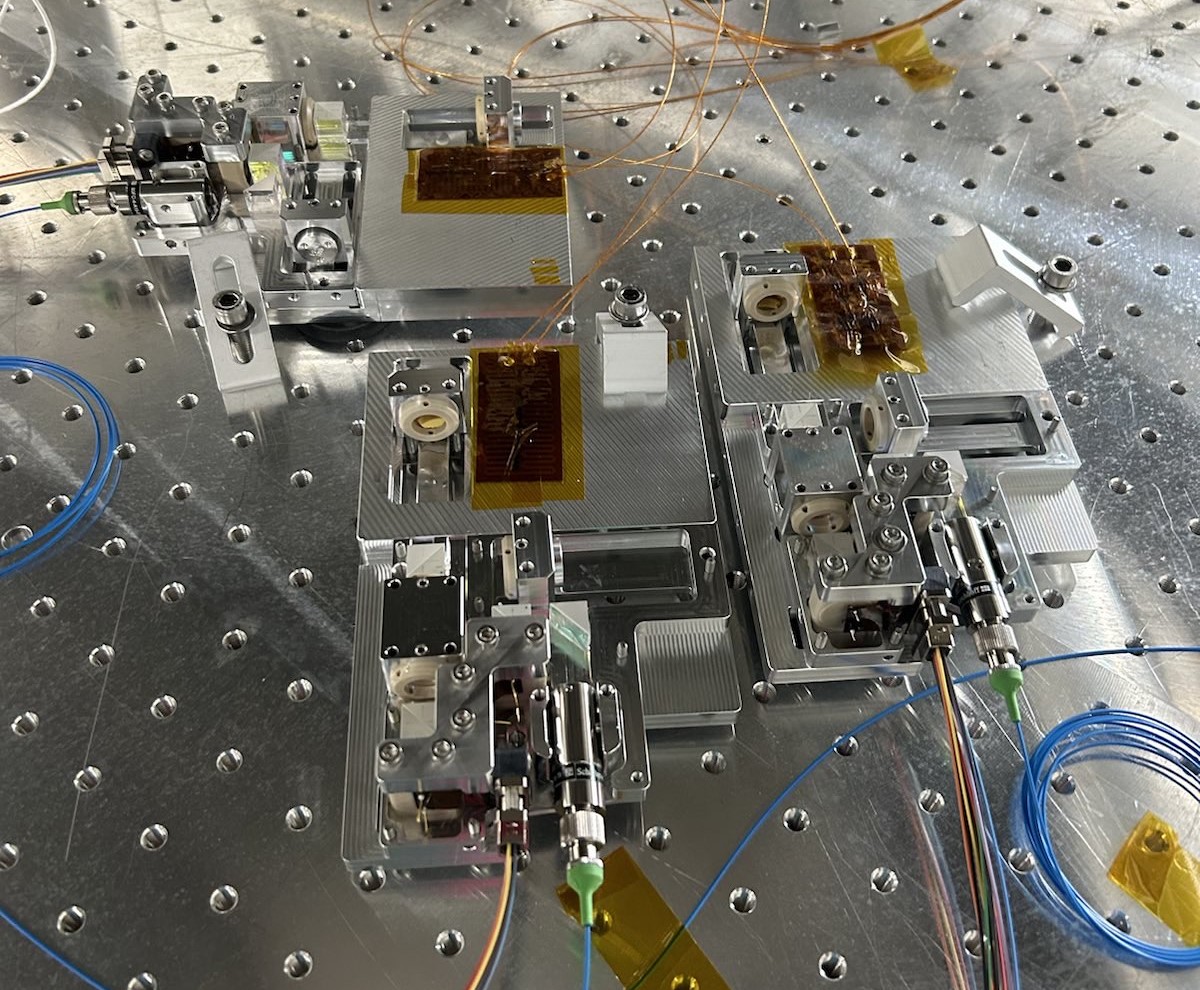}
    \caption{Three HoQIs set up for the static measurement. The orange coloured shapes attached to the extension plates are the resistors used to calibrate the HoQIs. HoQIs are on rubber feet with clamps to the bench to reduce the impact of table motion.}
    \label{fig:HoQIsetup3}
\end{figure}

Three HoQIs were set up as shown in Figure \ref{fig:HoQIsetup3}.
Care was taken to ensure that the cables and fibres had sufficient strain relief. 
Rubber feet and clamping of the baseplates provided some ground motion isolation, which is useful as the HoQIs were placed on a non-isolated bench. 
The advantage of using multiple units in these static measurements is that coherent subtraction can be done between them. 
This is done by the function `mccs2'  which is a multi-channel coherent subtraction \cite{mccs}.
The reduction in noise for HoQIs using mccs is primarily the subtraction of laser frequency noise, as all HoQIs use the same laser source. 
This method does not subtract the majority of seismic noise as the coherence length of seismic motion is small. 
The HoQIs were aligned so that the arm lengths were matched to within around 1mm to further reduce the impact of frequency noise. 

The signal and optics chain is almost identical to the one shown in Figure \ref{fig:GenSigOp}, however, for these tests the Virgo data acquisition system was used.
No ellipse fitting process was used as the motion was so small that no reasonable fraction of the ellipse was visible to fit to.
Instead, the photodiodes were calibrated so that any differences in gain or offset between them were corrected for. 
This was done using a resistor attached to the extension plate, which can be seen in Figure \ref{fig:HoQIsetup3}, which heated the baseplate and `drove' the HoQI through a full ellipse.
This allowed us to do calibration whilst the HoQIs were in vacuum so the photodiodes were under the same conditions as during the static measurements. 
The gains and offsets could then be extracted and used to normalise the photodiodes.

An NKT photonics Koheras laser provided the light source and this was taken in vacuum using a Diamond vacuum fibre feedthrough as mentioned in Section \ref{sec:optics}.
This was split in-vacuum using an OZ optics 1:8 fibre splitter, allowing fewer feedthroughs whilst maintaining a consistent laser source for frequency noise subtraction.

\begin{figure}
    \centering
    \includegraphics[scale=0.4]{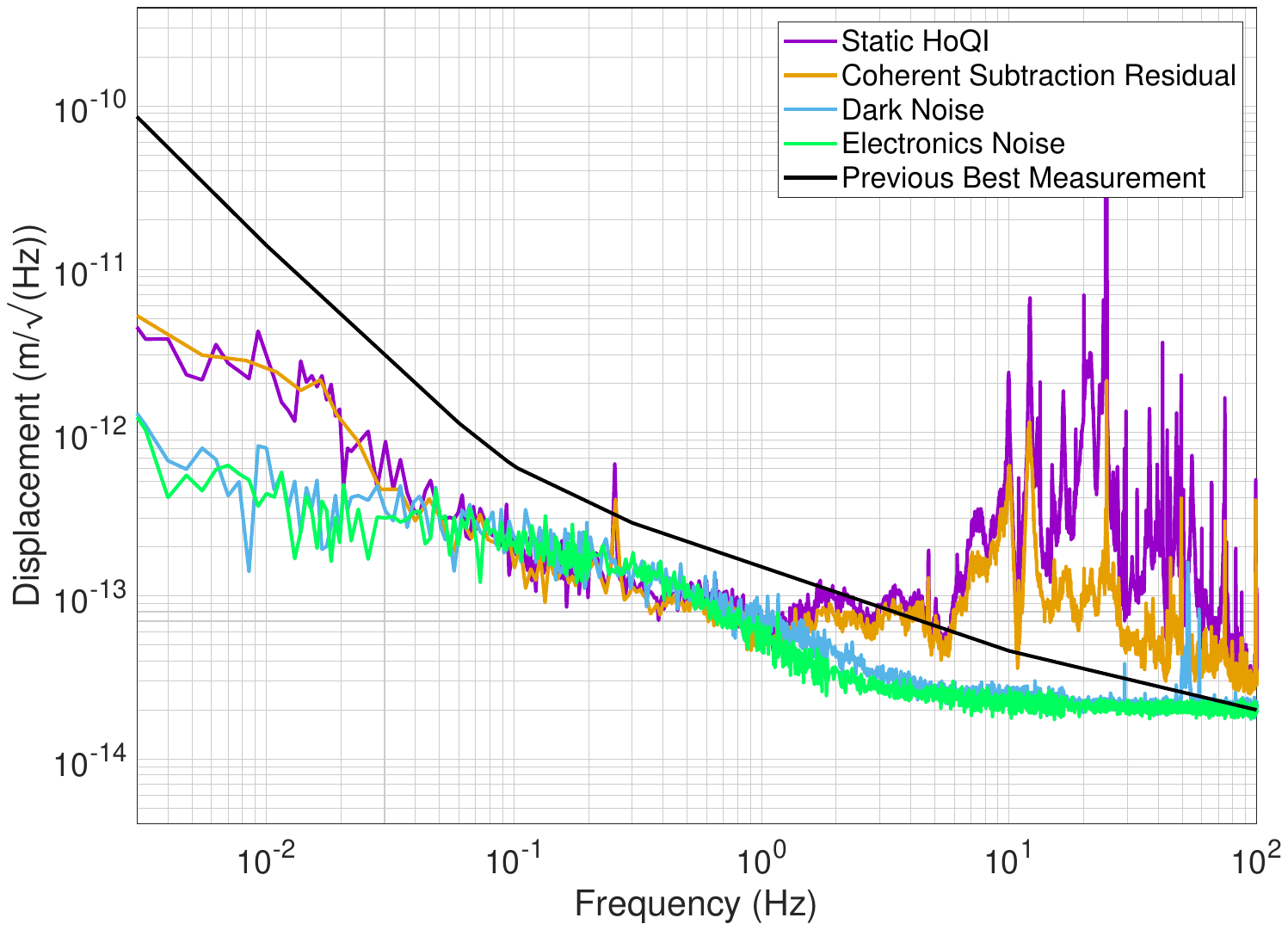}
    \caption{Static, in-vacuum results for one HoQI alongside the measured electronics and dark noise and the previous best measurement, which is based on results from Cooper 2018 \cite{Cooper2018}.}
    \label{fig:StaticNoiseBudget}
\end{figure}

\begin{figure}
    \centering
    \includegraphics[scale=0.4]{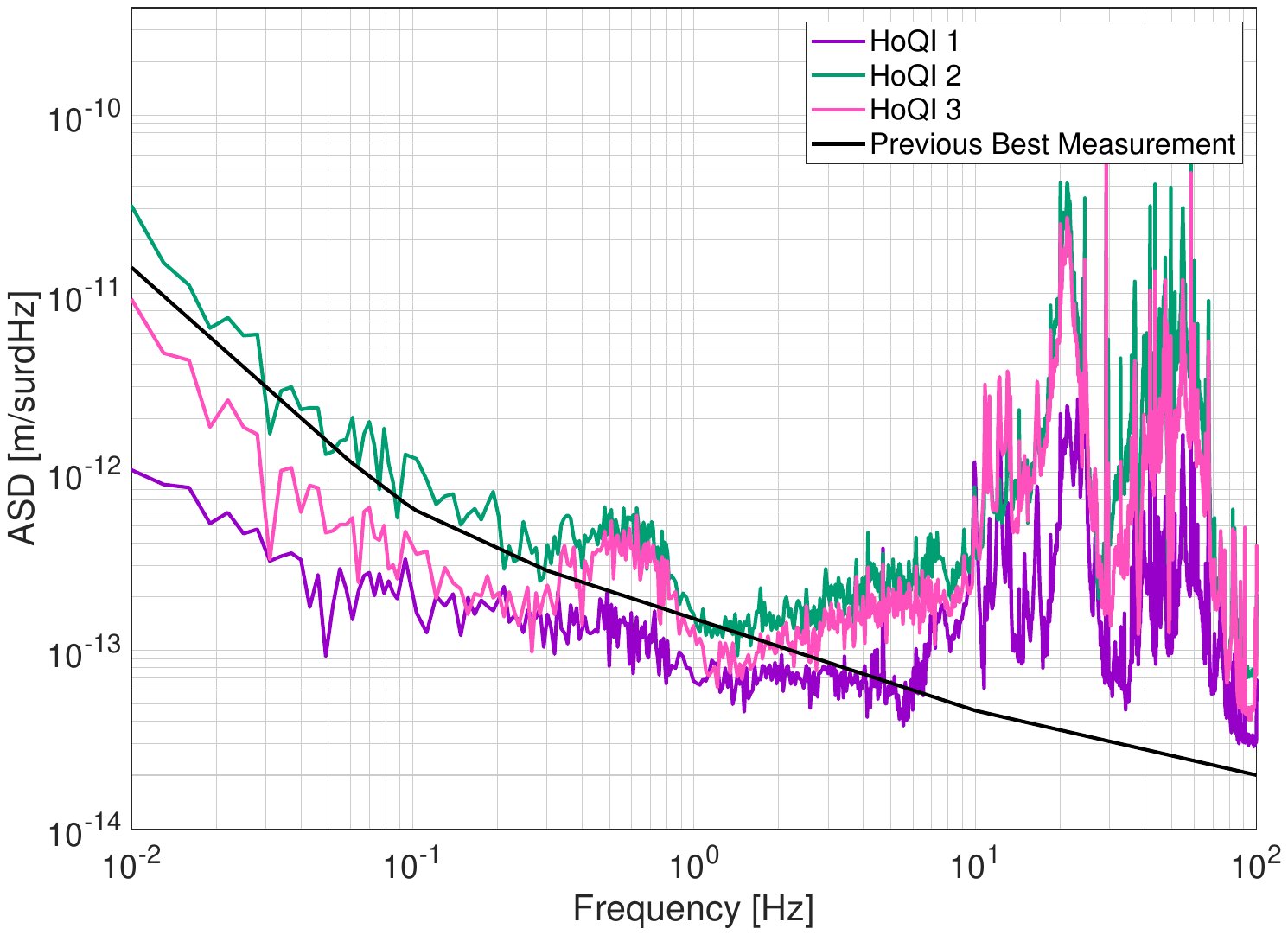}
    \caption{Static, in-vacuum results for all three HoQIs at one time.}
    \label{fig:Static3HoQIs}
\end{figure}

The results of these tests can be seen in Figure \ref{fig:StaticNoiseBudget} which gives the peak performance of one HoQI alongside some measured noise sources. 
The dark noise was measured by turning off the laser and the electronics measured by unplugging the HoQIs from the pre-amplifier. 
The coherent subtraction residual is the result of the mccs function which gets rid of the coherent signal between the HoQIs (the other two HoQI signals in the same time were subtracted from the signal shown).
In this case it should mainly subtract frequency noise. 
Figure \ref{fig:Static3HoQIs} shows the peak performance of all three HoQIs at one time.

The frequency noise was measured using HoQIs in a separate measurement. 
To do this two HoQIs were aligned in the static configuration but with mis-matched arm lengths of 30\,mm (measured using caliper), whilst one HoQI kept the matched arm length configuration (matching to around 1\,mm, the error on the measurement). 
We assume that by mis-matching the arm lengths the HoQI will be dominated by frequency noise. 
By rearranging Equation \ref{eq:freqNoise} as in Equation \ref{eq:freqMeas} we can calculate the frequency noise in an arm-length-matched HoQI.
Here $\Delta f$ is the frequency noise and $\Delta L$ is the arm length mismatch, 1 indicates the arm length matched HoQI and 2 indicates the mis-matched arm length HoQI.

\begin{equation}\label{eq:freqMeas}
    \Delta f_1 = \Delta f_2 \frac{\Delta L_1}{\Delta L_2}
\end{equation}

\begin{figure}
    \centering
    \includegraphics[scale=0.4]{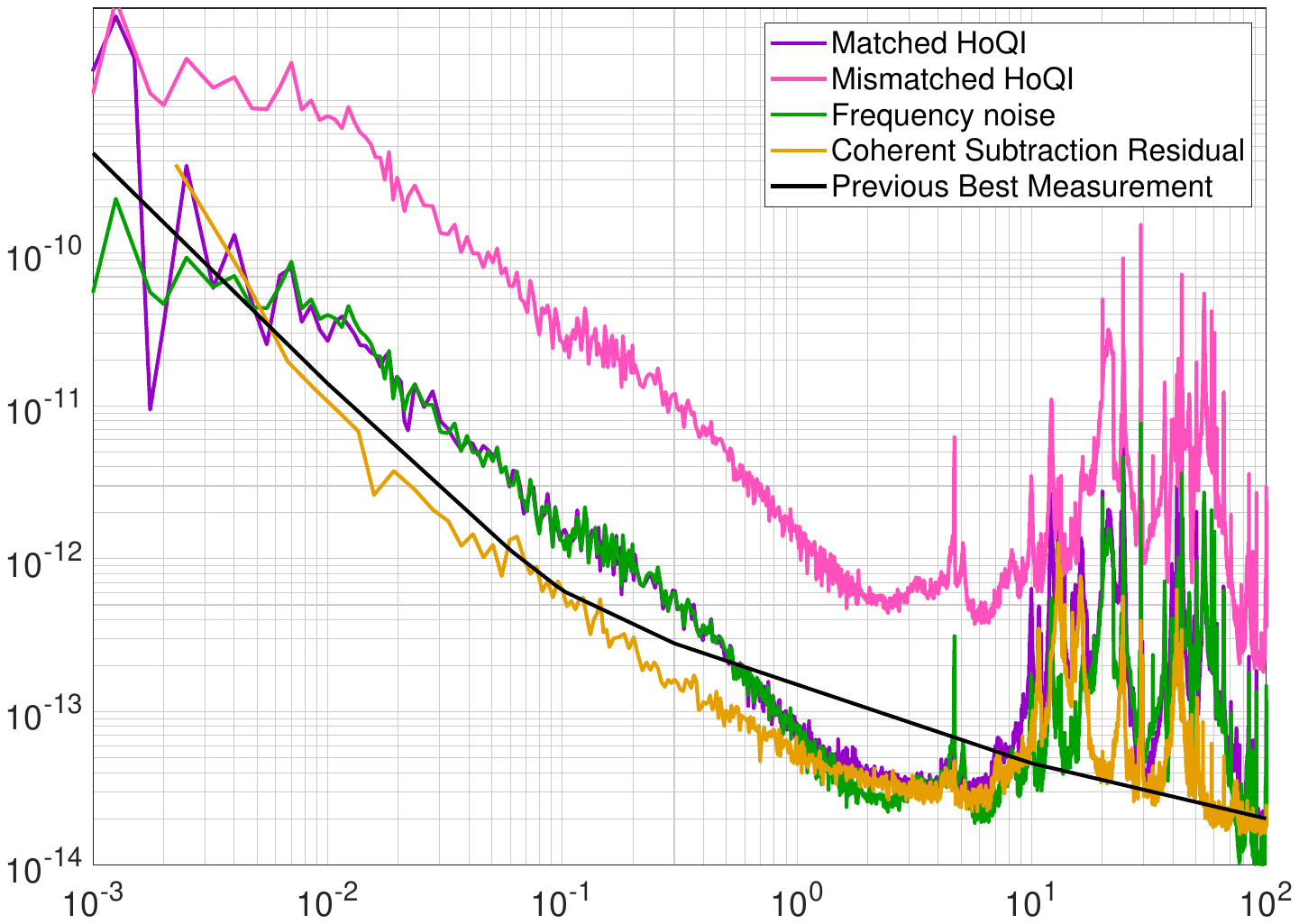}
    \caption{Frequency noise measurement using static HoQIs with mismatched arm lengths.}
    \label{fig:FreqNoise}
\end{figure}

The results of this measurement are given in Figure \ref{fig:FreqNoise}. 
The value of $\Delta L_1/\Delta L_2$ was adjusted to account for errors in the measured values and a `good' time period for the mis-matched HoQI spectra was chosen. 

The static performance of the upgraded HoQIs in vacuum is shown to be better across most of the bandwidth than the previous peak performance. 
All three HoQIs showed performance around the level indicated in Figure \ref{fig:StaticNoiseBudget} at some times but irregularities in the environment meant that the peak performance of each fell in different time intervals. 
The peak sensitivity is \num{2.5e-14}\,m/$\sqrt{\rm{Hz}}$ at 100\,Hz and at 10\,mHz it is \num{2.3e-12}\,m/$\sqrt{\rm{Hz}}$. 
The spectrum indicates a noise floor of around 3-\num{4e-14}\,m/$\sqrt{\rm{Hz}}$ above 1\,Hz.  
The dark and electronics noise could not be measured simultaneously with the noise performance of the HoQI.
From Figure \ref{fig:StaticNoiseBudget} it is assumed that dark noise is limiting between 30m\,Hz-1\,Hz and that it was slightly lower in the static measurement presented here.

Figure \ref{fig:FreqNoise} indicates that we are limited by frequency noise.
The residual of the coherent subtraction in Figure \ref{fig:StaticNoiseBudget} shows very little subtraction below 5Hz and is below the measured frequency noise, suggesting the arm lengths in this HoQI must be better matched than in the frequency noise measurement (within 1\,mm).
This improved arm-length-matching is further demonstrated by the difference in performance between the frequency noise `matched' spectrum in Figure \ref{fig:FreqNoise} and the static performance in Figure \ref{fig:StaticNoiseBudget}.
As we chose a quiet time period for the mismatched HoQIs we assume there is not excess noise present in them during this measurement. 

Above 5\,Hz the noisy lab environment creates many peaks in the spectrum, some of which can be reduced using the coherent subtraction, in particular the 20\,Hz peak which is visible across all spectra. 
This indicates that some motion in the lab is coupling to frequency noise; We suspect this comes from the laser or other optics on the laser bench. 

The electronics and dark noise were below the static noise above 1.5\,Hz, but these noises are limiting between 30\,mHz-1.5\,Hz. 
Without the noisy lab environment these noises will likely set the noise floor. 
We found that the dark noise was impacted significantly by strain relief in the cables. 
It is expected that laser noise (frequency, intensity) is the dominant source of noise below 30\,mHz, but this is difficult to measure without significant additions to the setup.

\begin{figure}
    \centering
    \includegraphics[scale=0.4]{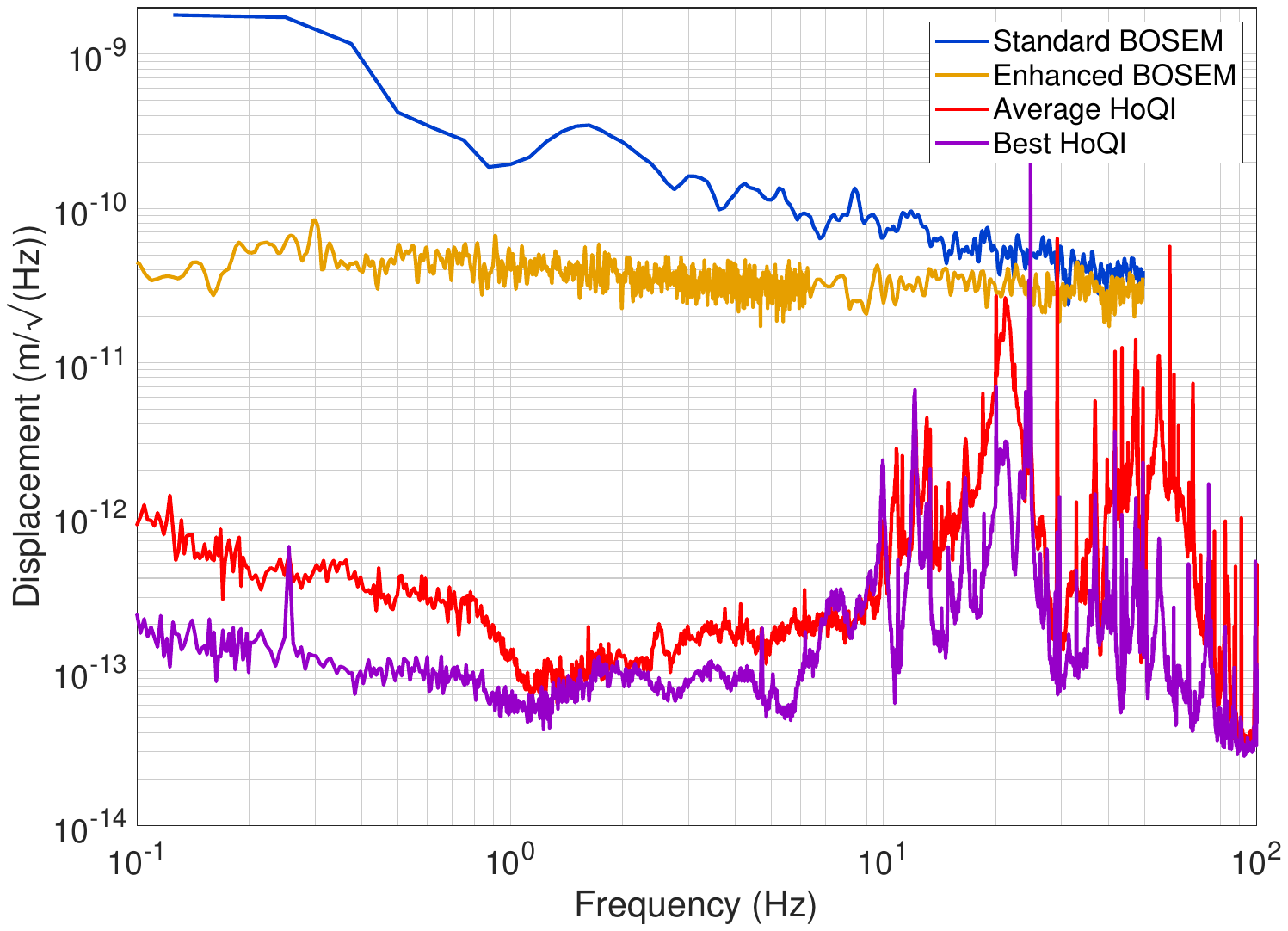}
    \caption{Standard and enhanced BOSEM measurements \cite{Cooper2023} compared to the best and average static measurements from HoQI.}
    \label{fig:BOSEMcompare}
\end{figure}

Figure \ref{fig:BOSEMcompare} shows the comparison of the most commonly used sensor in GW detectors, the BOSEMs, with HoQIs.
This takes the standard and enhanced BOSEM results from \cite{Cooper2023} and compares it to the best measurement of HoQI in these static tests and a measurement which is at the average static HoQI performance. 
This shows that HoQI is quieter than BOSEMs across the entire frequency range by multiple orders of magnitude. 

\section{Mounting on Suspensions}
\label{sec:suspensions}

Due to its increased sensitivity HoQI can be used further down the suspension chain (where the motion is smaller) compared to other displacement sensors. 
This gives greater control authority of and reduces the motion of the optic without injecting sensing noise. 
The benefits of using HoQIs for damping the BBSS have been modelled in \cite{Dongen2022} demonstrating improved damping of suspension modes in the optics.

The performance of singular HoQIs has been demonstrated, however, for practical application in GW detectors multiple HoQIs must work together.
The following experiment  employs multiple HoQIs to measure a mass in a multi-stage suspension, situated closer to the optic than the BOSEMs, demonstrating a crucial step toward the utilisation of HoQIs for local control in gravitational wave detectors.
This experiment was done on a non-isolated table but to demonstrate system identification and control using the HoQIs the suspension needs to be installed in the interferometer and the global controls functional, which is beyond the scope of this paper. 
The experiment will be explained in this section with results discussed in Section \ref{sec:discussion}. 

The AEI 10\,m prototype is a 10\,m long Fabry-Perot Michelson interferometer located at the Albert Einstein Institute (AEI) in Hannover.
The interferometer optics are suspended in vacuum with multi-stage suspension systems on seismically isolated tables, AEI's seismic attenuation system (AEI-SAS), which use both active and passive isolation \cite{Kirchhoff2020}.
The control and data system uses the software developed for LIGO and encompases all aspects of the control, monitoring and data collection for the entire interferometer \cite{Bork2010}.
The prototype serves as a testbed for current and upcoming technologies and similarities in suspension and control systems between the AEI 10m prototype, and gravitational wave observatories (specifically LIGO) make it an excellent option to test HoQIs on multi-stage suspensions.

LIGO's BBSS intermediate mass were designed to have space around the BOSEMs to mount more sensors, such as HoQIs \cite{BBSS}. 
This configuration makes it an ideal candidate for modeling damping using HoQIs, as in \cite{Dongen2022}.
The AEI beamsplitter is a comparable to the BBSS but it is already operational, making it a suitable option for this experiment. 
However, unlike LIGO suspensions, accesible and verified models of the AEI interferometer do not exist, so comparisons will be made to the BBSS results.

\begin{figure}
    \centering
    \includegraphics[scale=0.5]{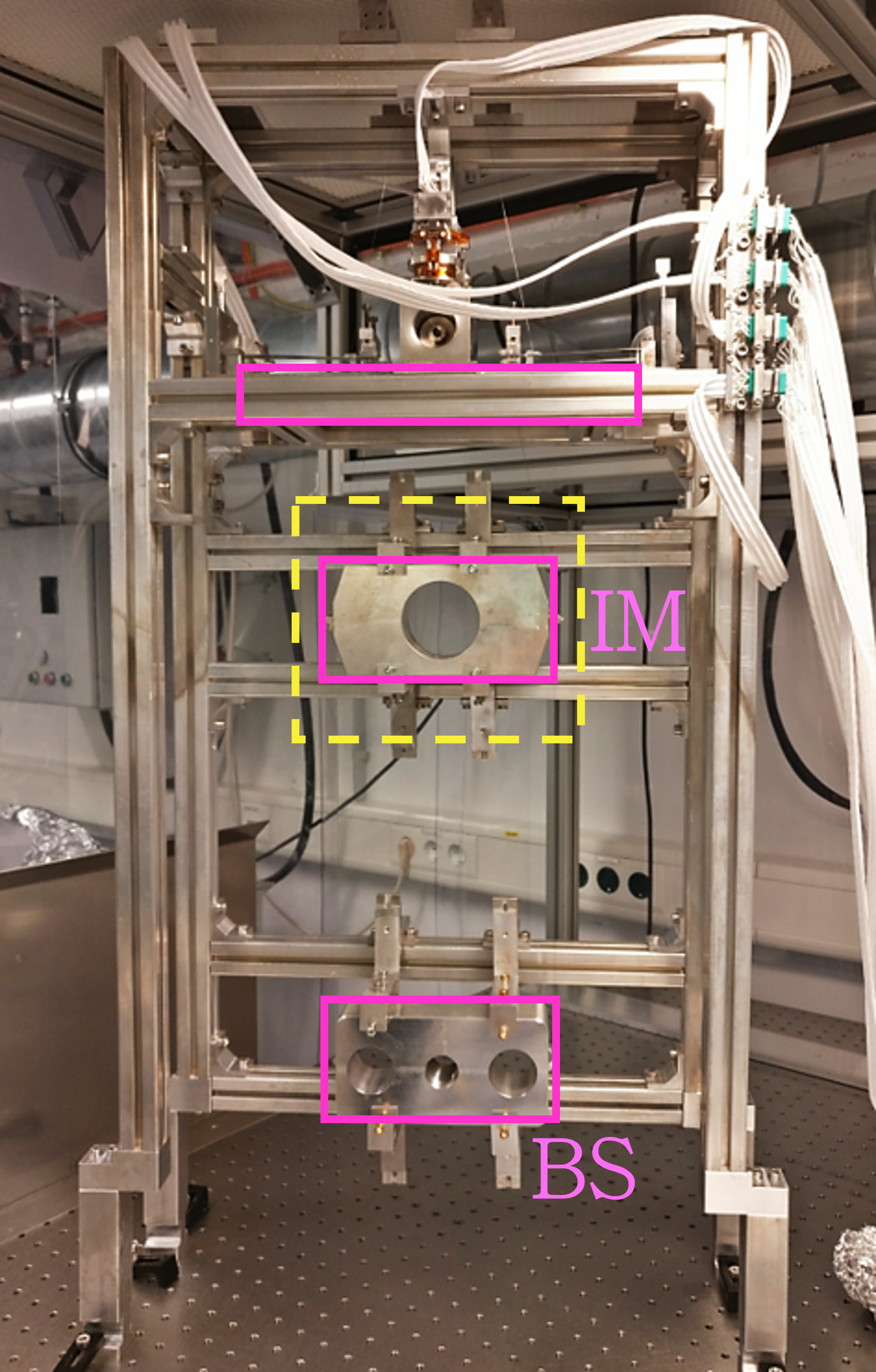}
    \caption{A picture showing the AEI beamsplitter triple suspension. The three masses are indicated by solid pink boxes, IM is the intermediate mass and BS is the beamsplitter (in this photo an aluminium dummy mass with the same mass and inertia is in place of the optic) and the position of the tablecloth on which the HoQIs were mounted (on the other side of the suspension) is given by the dashed yellow line box.}
    \label{fig:BSpic}
\end{figure}

\begin{figure}
    \centering
    \includegraphics[scale=0.6]{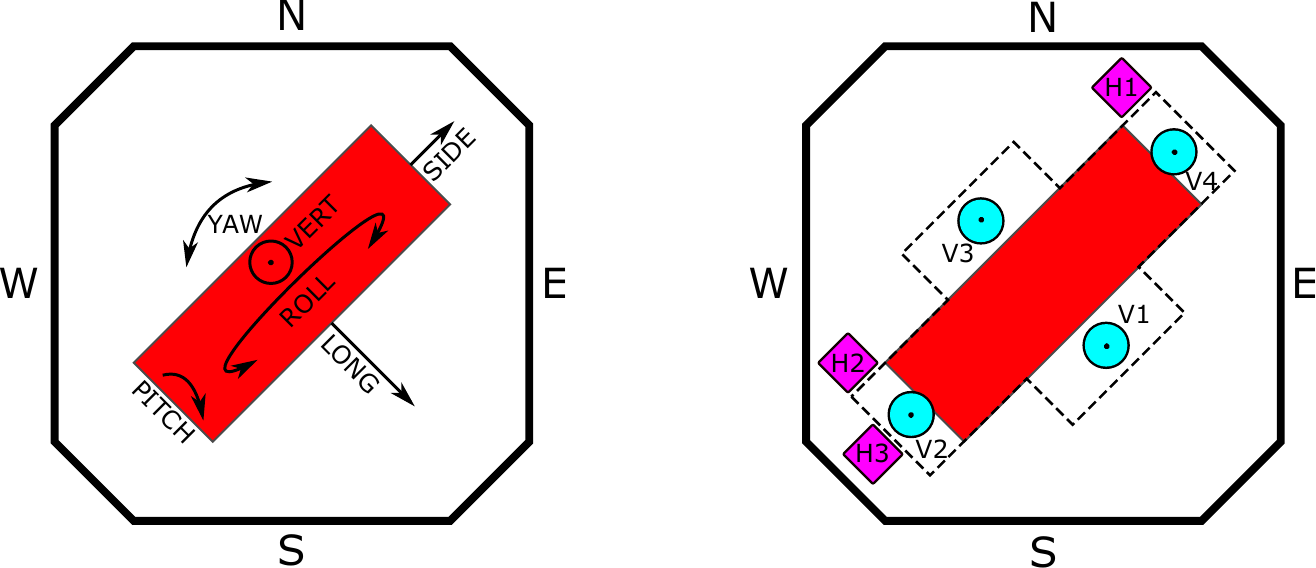}
    \caption{Schematic of the top down view of the AEI beamsplitter suspension top mass, on the left the degrees of freedom labelled and on the right the positions of the BOSEMs, with H being a horizontal BOSEM and V being a vertical BOSEM.}
    \label{fig:TopMass}
\end{figure}

\begin{figure}
    \centering
    \includegraphics[scale=0.8]{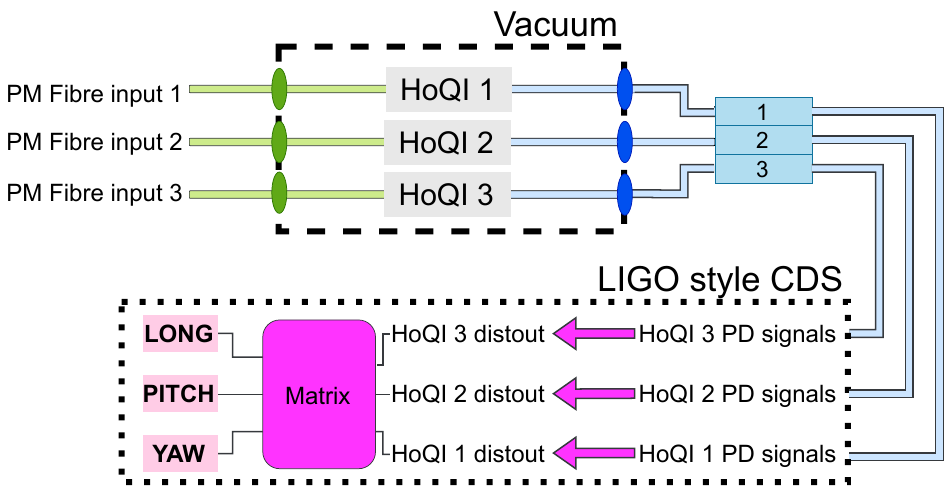}
    \caption{Figure showing the signal and optics chain used at the AEI 10m prototype. Three HoQIs are used and must function together to give a useful output. An optomechanical interface will be provided by a retroreflector attached to the intermediate mass for the beamsplitter suspension. The data acquisition and processing is done using a LIGO-style control and data system (CDS) and this is used to process the three HoQI outputs to degrees of freedom. }
    \label{fig:AEISigOp}
\end{figure}

The AEI 10\,m prototype beamsplitter is a 1.5\,kg optic mounted on a triple suspension, pictured in Figure \ref{fig:BSpic}. 
Seven BOSEMS are used to measure and damp motion of the top mass.
Their position and relation to the interferometer degrees of freedom can be seen in Figure \ref{fig:TopMass}. 
At the time of this experiment the beamsplitter was not in vacuum or mounted on the AEI-SAS, instead on an optical bench inside a clean tent nearby. 
However, to ease transition and get consistent results all fibre optics and cables used here will not change after the move. 
Ground motion in the prototype facility is measured using an STS2 seismometer aligned to the axes of the arms.
The signal and optics chain for this experiment is in Figure \ref{fig:AEISigOp}.

\begin{figure}
    \centering
    \includegraphics[scale=0.7]{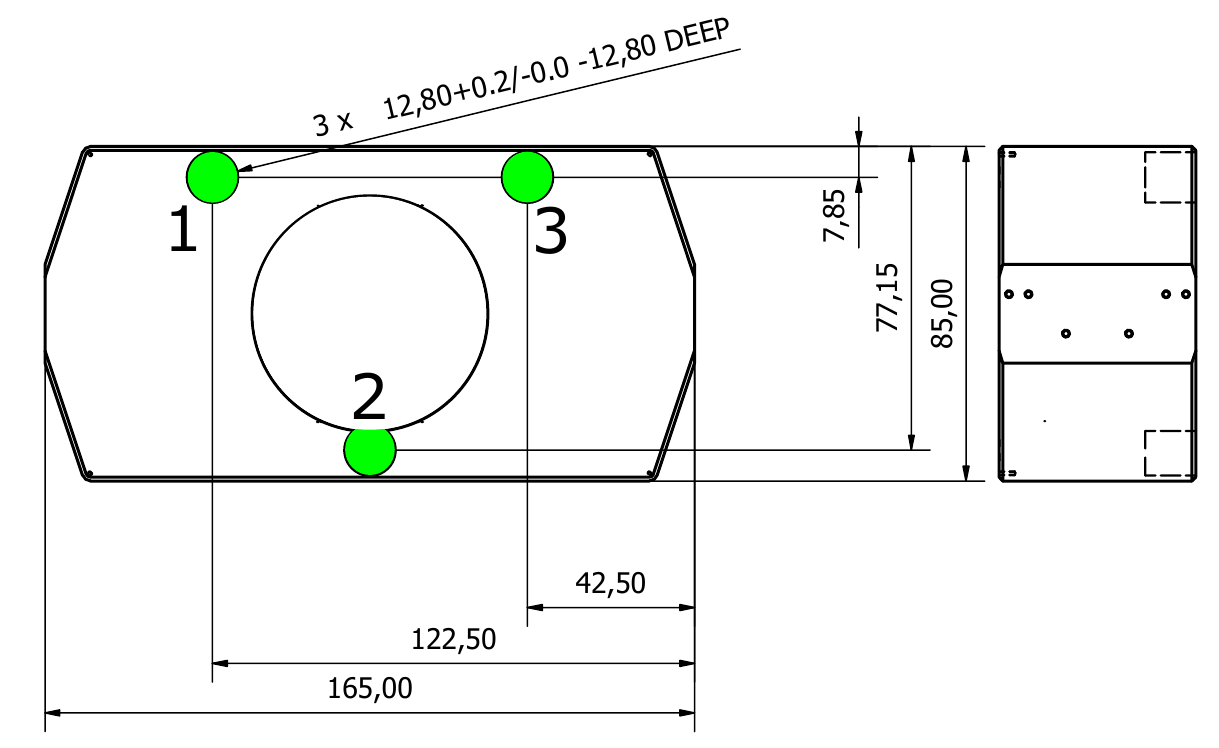}
    \caption{Schematic of the IM with the retroreflector holes in it in green and HoQI numbers labelled below. All numbers labelled are in units of mm.}
    \label{fig:IMSchematic}
\end{figure}

Three retroreflector HoQIs were mounted on the intermediate mass (IM) tablecloth, indicated by the yellow dash box in Figure \ref{fig:BSpic}. 
They can be seen mounted onto the suspension in Figure \ref{fig:AEImounting}.
The HoQIs will measure differential displacement between the frame and the IM with three hollow, gold retroreflectors glued inside notches in the IM as the optomechanical interface. 
Figure \ref{fig:IMSchematic} is a drawing of the IM with the positions of the retroreflectors indicated.
From this we can obtain an ideal sensing matrix for longitudinal (long), pitch and yaw: the degrees of freedom which will be probed, given in Table \ref{tbl:sensinghoqi}. 

\begin{table}[h]
    \centering
    \caption{Idealised sensing matrix for HoQIs 1, 2 and 3 to the long, pitch and yaw degrees of freedom}
    \label{tbl:sensinghoqi}
    \begin{tabular}{|c|c|c|c|}
    \hline
                   & \textbf{HoQI 1} & \textbf{HoQI 2} & \textbf{HoQI 3} \\ \hline
    \textbf{Long}  & 0.25            & 0.5             & 0.25            \\ \hline
    \textbf{Pitch} & -7.21           & 14.42           & -7.21           \\ \hline
    \textbf{Yaw}   & -12.5           & 0               & 12.5            \\ \hline
    \end{tabular}
\end{table}

\begin{figure}
    \centering
    \includegraphics[scale=0.45]{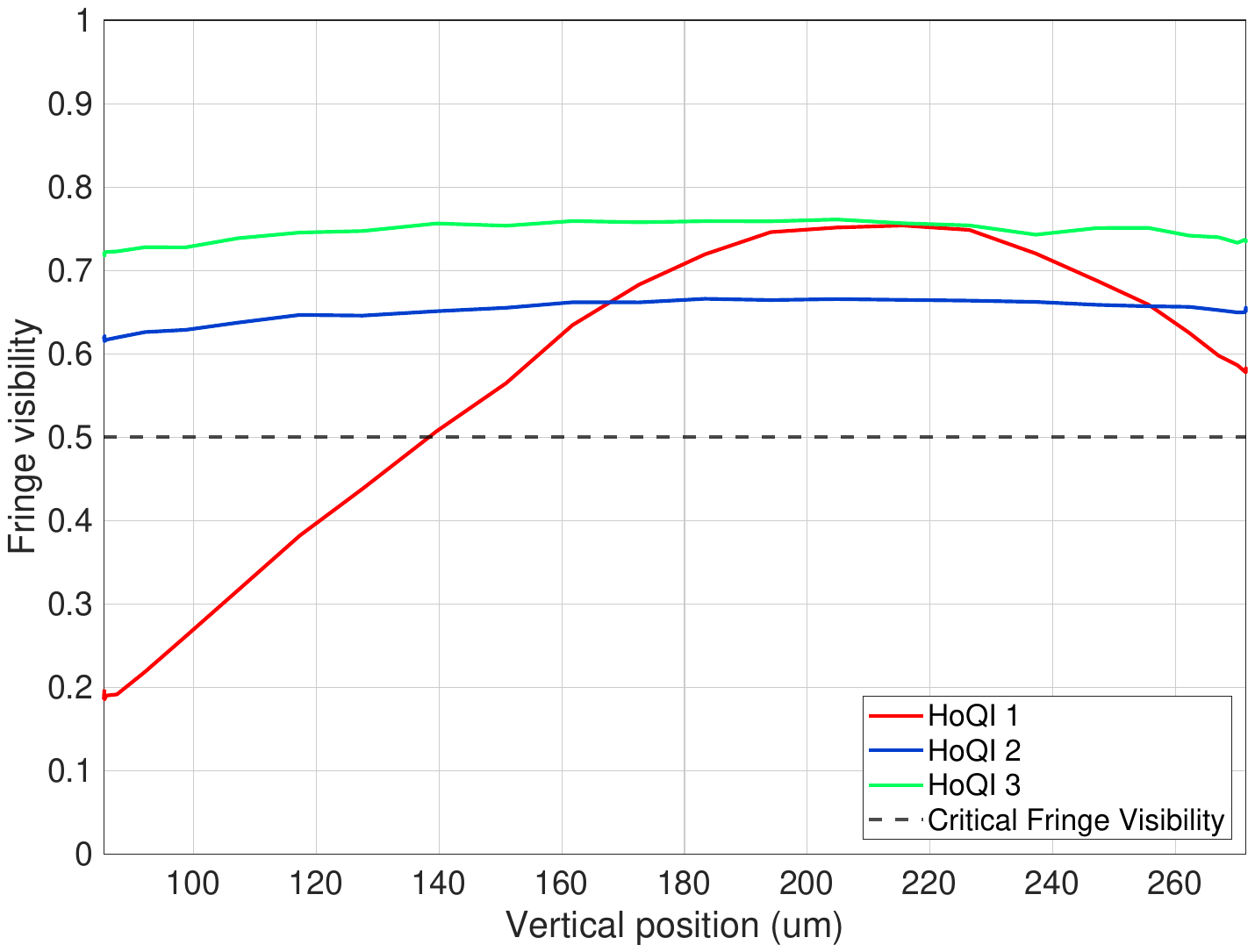}
    \caption{The change in fringe visibility as a function of vertical misalignment. The intermediate mass was driven vertically (a degree of freedom not measured using the HoQIs) using the top stage BOSEMS and the fringe visibility was observed.}
    \label{fig:FVmeas}
\end{figure}

The mounting system is as described in Section \ref{sec:mechanics} and careful design ensures that the arm lengths of the HoQI can be matched.
A key step to ensure HoQI measurements are not subject to large non-linearities is the internal alignment of optics (waveplates, fibre collimator and arm optics) which must be done prior to mounting on the suspension. 
All HoQIs showed fringe visibility around 0.7 after mounting and external alignment to the IM, higher than the critical fringe visbility value. 
Misalignment in alternate degrees of freedom may degrade this value. 
As discussed in Section \ref{sec:optics}, however, it is vital for performance that the fringe visibility remains over this critical value throughout the possible range of motion. 
To test the robustness of the fringe visibility in the AEI suspensions, the BOSEMs were used to drive the IM vertically orthogonal to HoQI sensing.
The IM was moved through the maximum drive of the BOSEMs which is a larger range than could be expected during low-noise operation of the interferometer and the fringe visibility was tracked across this range of motion. 
The results of this are shown in Figure \ref{fig:FVmeas}.

\begin{figure}
    \centering
    \includegraphics[scale=0.45]{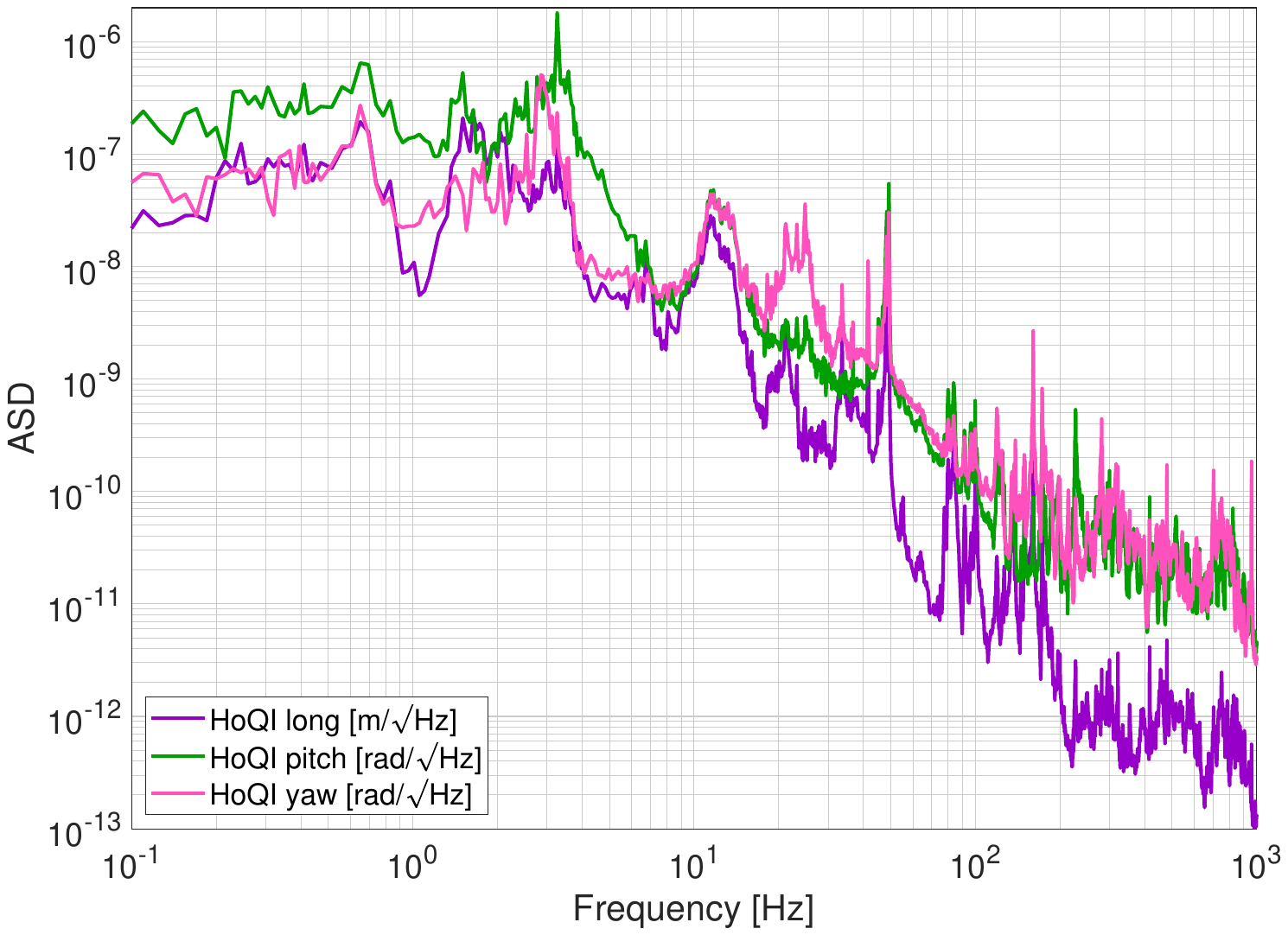}
    \caption{The ellipse fitted HoQI measurements of the intermediate mass of a triple suspension in all three measureable degrees of freedom: Long, pitch and yaw.}
    \label{fig:HoQIDofs}
\end{figure}

\begin{figure}
    \centering
    \includegraphics[scale=0.35]{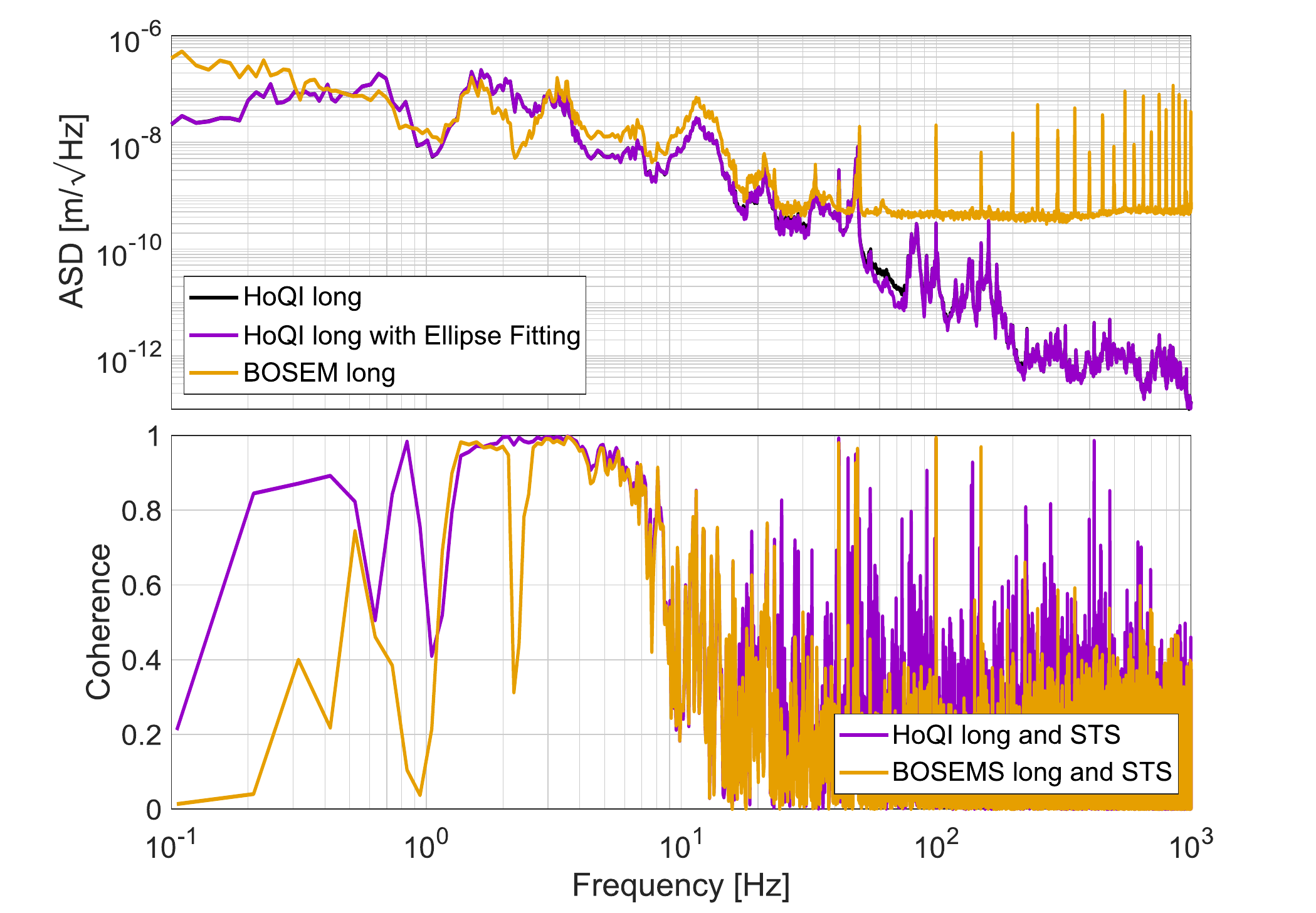}
    \caption{a) Figure showing the long degree of freedom measurement using the BOSEMS on the top mass and the HoQIs on the intermediate mass. The HoQI measurement is shown both with and without ellipse fitting, done here in post processing. b) The coherence of the STS2 seismometer with both HoQI long (ellipse fitted) measurement and BOSEMS long measurement. The STS measurement was fitted to an angle as the seismometer and suspension did not have the same coordinate direction.}
    \label{fig:LongCompare}
\end{figure}

To ensure non-linearities would not affect our results, multiple precautions were taken as mentioned in Section \ref{sec:optics}. 
As the beamsplitter was not on an inertial platform, ground, and therefore cage, motion will be high. 
To minimise this effect BOSEM top mass damping was employed and we ensured the HoQIs were well aligned, as discussed above.
We can use the ellipse fitting process as both a tool for correcting for non-linearities in the data and monitoring how much non-linearity there is present in a measurement. 
This tool has been shown to reduce non-linearity to a suitable level, as demonstrated in \cite{Cooper2020}.
Figure \ref{fig:HoQIDofs} gives a degree-of-freedom comparison for ellipse fitted data and Figure \ref{fig:LongCompare}a shows the BOSEM output compared to HoQI raw measurements and HoQI ellipse fitted measurements.

The cage motion of the beamsplitter as a result of ground motion will dominate the sensor readouts. 
Therefore the STS seismometer can be used as an indicator for how well each sensor is performing. 
The coherence between the seismometer and the HoQI and BOSEMs is shown in Figure \ref{fig:LongCompare}b.

\section{Discussion}
\label{sec:discussion}
Figure \ref{fig:HoQIDofs} shows the ASD of the HoQIs in all three measured degrees of freedom. 
A resonance can be seen in all degrees of freedom at 11\,Hz.
Geophone measurements of the transfer functions between the floor, table and BOSEMs showed this peak was a combination of the table and suspension resonances.

Figure \ref{fig:LongCompare}a) shows the BOSEMs and HoQI measurements of their respective masses in the long degree of freedom.
Above 50\,Hz the BOSEMs measurement is consistent with a noise floor of $3 \times 10^{-10} \, \rm{m} \, \sqrt{\rm{Hz}}$, which is above the noise of an enhanced BOSEM but matches the standard BOSEM noise.
The HoQIs have a significantly lower noise floor, as seen in Figure \ref{fig:StaticNoiseBudget}, allowing them to detect motion of the intermediate mass with significant SNR.
The HoQI long ASD above 50\,Hz is still highly structured, implying we are measuring real motion instead of noise. 
Both the HoQI and BOSEMs see the table and suspension resonance around 11\,Hz.

Ellipse fitting corrects for and quantifies how much non-linearity there is in the system. 
Good alignment of the HoQIs after installation is evidenced by the close match between the ellipse fitted and non-ellipse fitted traces over the majority of the frequency range.
The ellipse fitting corrects for the most non-linearities between 10-100\,Hz.
When motion is high, indicated by a sharp peak, the non-linearity effect causes up- and down-conversion of the signal. 
This is explained and demonstrated further in Section \ref{sec:optics} and Section 3 of \cite{Cooper2020}. 

Due to the fact the suspension was sat on an unisolated optical bench, the motion measured by both HoQI and BOSEMs is predominantly `cage motion' which is driven by ground motion (the `cage' is the structure on which the sensors are mounted surrounding the pendulum). 
This explains the similarities between the two sensors and further evidence can be seen for this in Figure \ref{fig:LongCompare}b).

Figure \ref{fig:LongCompare}b) shows the coherence between the HoQI/BOSEM and the STS2 seismometer. 
Above 10\,Hz the ground motion measurements and the HoQI and BOSEM measurements are incoherent. 
This is due to both the distance between the sensors (a few metres) and the table and suspension resonance at 11\,Hz. 
At frequencies below 10\,Hz there is significantly more coherence, as expected. 
Dips in the coherence, for example in the BOSEM at 2.3\,Hz, arise from resonances which cause cross coupling in other degrees of freedom. 
The HoQI shows better coherence with the STS2, indicating that there is a better signal to noise ratio than the BOSEMs. 
We expect the measurement of the ground from the STS and the measurement of the suspension to test mass to be completely coherent at these frequencies, hence incoherence indicates increased noise. 
Below 1\,Hz increasing noise in the sensors reduces the coherence to ground motion (lower signal to noise ratio).
Coherence is maintained better in HoQI in comparison to the BOSEMs further demonstrating its greater low frequency performance and operation in air.

The change in fringe visibility as a function of misalignment can be seen in Figure \ref{fig:FVmeas}. 
For installation in suspensions, the fringe visibility should be robust across the range of motion expected (even if not measured) and should remain above the critical fringe visibility.
There is good robustness to misalignment in all HoQIs across a 140\,$\mu$m range. 
Two out of the three HoQIs (HoQI 2 and HoQI 3) demonstrate good collimation and alignment by maintaining fringe visibility above the critical value of 0.5 across the full 200\,$\mu$m. 
HoQI 1 shows a reduction in fringe visibility below the critical value over the full range of motion tested. 
We believe this is due to a modification in the collimation of the beam, which affects the spot size on the optics.
The beam needs to be as large as possible at the photodiodes for good fringe visibility throughout misalignment, as shown in equation \ref{eq:FVmis}. 
To improve the beam size the collimator would need to be removed and adjusted, a disruptive process to the functional system hence why this wasn't done. 
The fringe visibility is robust across a range of motion exceeding expected suspension values provided the collimation of the light is optimal within the sensor.


\section{Conclusion}
\label{sec:conclusion}
This paper has described the design and demonstrated performance and practical use of HoQIs. 
The design of HoQI has been driven by constraints set for different use cases whilst maintaining good resolution. 
The fringe visibility and non-linearities have been found to have an impact on the performance of HoQI, but this effect can be made insignificant by using high quality optics and optical chain, good alignment of the system and ellipse fitting. 
We have demonstrated that HoQI is a reproducible and integratable sensor by building multiple functional units and getting them to perform in multiple ways and environements. 
The peak static, in-vacuum sensitivity of the HoQIs is \num{2.5e-14}\,m/$\sqrt{\rm{Hz}}$ at 100\,Hz and at 10\,mHz it is \num{2.3e-12}\,m/$\sqrt{\rm{Hz}}$ with a noise floor of around 3-\num{4e-14}\,m/$\sqrt{\rm{Hz}}$ above 1\,Hz.
In contrast to the BOSEMs, the HoQI noise floor was never reached during measurements of the AEI beamsplitter suspension intermediate mass. 
The HoQI showed better coherence with the ground motion measurement than the BOSEMs at low frequencies demonstrating a better signal-to-noise ratio.
The fringe visibility as a function of misalignment was measured and with good collimation the robustness was good, with no significant drop in fringe visility over 200\,$\mu$m. 

HoQI has potential to be integrated into current and future gravitational wave detectors as the readout of other sensors (such as the CRS) and to measure and damp suspension motion (such as on the BBSS). 
CRS HoQIs have already undergone preliminary testing and this is expected to continue.
For integration onto suspensions the AEI 10\,m beamsplitter first needs to be placed in vacuum on the AEI-SAS.
The BOSEM and HoQI measurements can then be done with lower input motion and the HoQI actuation can be tested to see how the suspension performance changes. 
As modelled in \cite{Dongen2022}, this should help to damp the suspension resonances and therefore reduce optic motion, improving the interferometer performance.




\section{Acknowledgements}
\label{sec:Acknowledgement}
This project has received funding from the European Research Council (ERC) under the European Union's Horizon 2020 research and innovation programme (grant agreement No. 865816). The authors acknowledge support from the International Max Planck Research School (IMPRS) on Gravitational Wave Astronomy and the Deutsche Forschungsgemeinschaft (DFG, German Research Foundation) under Germany’s Excellence Strategy—EXC-2123 QuantumFrontiers—390837967. The authors acknowledge the support of the Institute for Gravitational Wave Astronomy at the University of Birmingham, STFC grants `Astrophysics at the University of Birmingham’ grant ST/S000305/1 and 'The A+ upgrade: Expanding the Advanced LIGO Horizon` ST/S00243X/1. 


\end{document}